\documentclass[intlimits,twoside,a4paper]{article}

\usepackage[cp1251]{inputenc}

\usepackage[eqsecnum]{cmpj3}

\usepackage{bm}

\issue{2023}{26}{2}{23702}
\doinumber{10.5488/CMP.26.23702}

\title[Collective excitations in three-band superconductors]%
{Collective excitations in three-band superconductors}
\author[K. V. Grigorishin]{K. V. Grigorishin\orcid{0000-0002-6195-7944}\thanks{Corresponding author: \email{konst.phys@gmail.com}.}}
\address{Bogolyubov Institute for Theoretical Physics of the National Academy of Sciences of
Ukraine, 14-b~Metrolohichna str., 03143 Kyiv, Ukraine}


\date{Received September 01, 2022, in final form October 30, 2022}

\begin{document}

\maketitle

\begin{abstract}
We investigate equilibrium states, magnetic response and the normal oscillations of internal degrees of freedom (Higgs modes and Goldstone modes) of three-band superconductors accounting the terms of both internal proximity effect and the ``drag'' effect (intergradient interaction) in the Lagrangian. Both the Goldstone mode and the Higgs mode are split into three branches each: common mode oscillations and two modes of anti-phase oscillations, which are analogous to the Leggett mode in two-band superconductors. It is demonstrated that the second and third branches are nonphysical, and they can be removed by special choice of coefficients at the ``drag'' terms in Lagrangian. As a result, three-band superconductors are characterized by only single coherence length. Spectrum of the common mode Higgs oscillations has been obtained. The magnetic penetration depth is determined with densities of superconducting electrons in each band, although the drag terms renormalize the carrier masses.

\printkeywords
%
\end{abstract}



\section{Introduction}\label{intro}

As well known, there is some analogy between particle physics and condensed matter. Thus, the nonrelativistic analog of the Higgs effect represents penetration of magnetic field in a superconductor. As a result of spontaneous broken gauge symmetry below $T_{c}$, the magnetic field gains the mass, the reciprocal value of which characterizes the penetration depth of the magnetic field in the superconductor. In the work~\cite{grig2} it is demonstrated that there are two types of collective excitations with the quasi-relativistic spectra in the single-band superconducting (SC) system: the Higgs mode $E^{2}=m_{\rm H}^{2}\upsilon^{4}+p^{2}\upsilon^{2}$, where $m_{\rm H}$ is the mass of a Higgs boson, so that $m_{\rm H}\upsilon^{2}=2|\Delta|$, and the Goldstone mode $E=p\upsilon$. The value $\upsilon=v_{\rm F}/\sqrt{3}$, where $v_{\rm F}$ is the Fermi velocity, plays the role of the speed of light, $|\Delta|$ is the energy gap in SC state. The Higgs mode is represented by oscillations of modulus of the Ginzburg-Landau order parameter (OP) $|\Psi(t,\mathbf{r})|$ and it can be presented as counterflows of SC and normal components so that $n_{\mathrm{s}}\mathbf{v}_{s}+n_{\mathrm{n}}\mathbf{v}_{n}=0$. This oscillation mode is unstable due to the decay into the above-condensate quasiparticles, since its energy is such that $E(q)\geqslant 2|\Delta|$. The Goldstone mode is represented by oscillations of the phase $\theta(t,\mathbf{r})$ of the OP $|\Psi|\re^{\ri\theta}$, which are the eddy currents $\mathrm{div}\mathbf{J}=0$ that are absorbed into the gauge field $A^{\mu}$ according to Anderson-Higgs mechanism. Thus, both Higgs mode and Goldstone mode are not accompanied by the charge density oscillations. At the same time, according to another model~\cite{ars}, Coulomb interaction ``pushes'' the frequency of the acoustic oscillations to the plasma frequency $\omega_{p}=4\piup ne^{2}/m$. Thus, the Goldstone mode becomes inherently unobservable since it turns to plasma oscillations. It should be noted that the Higgs and Goldstone bosons are typical of condensed matter. Thus, except superconductors, these bosons are observed in superfluid $^{3}$He-B and $^{3}$He-A~\cite{volovik}, although unlike the particle physics, the observed Higgs bosons are not fundamental: it comes as a composite object emerging in the fermionic vacuum.

The dynamics of multi-band superconductors is much more complicated than the dynamics of single-band superconductors due to the presence of several coupled OP $\Psi_{1},\Psi_{2},\ldots,\Psi_{n}$, i.e., multiband superconductors have a new property, such as the interband phase differences $\theta_{i}-\theta_{k}$. Two-band systems are the simplest but the most numerous class of multi-band superconductors. Their typical representatives are classical two-band superconductor magnesium diboride $\mathrm{MgB}_{2}$, nonmagnetic borocarbides $\mathrm{LuNi_{2}B_{2}C}$, $\mathrm{YNi_{2}B_{2}C}$ and some oxypnictide compounds~\cite{asker7}.  Two-band superconductor is understood as two single-band superconductors with the corresponding condensates of Cooper pairs $\Psi_{1}$ and $\Psi_{2}$ (so that densities of SC electrons are $n_{\mathrm{s}1}=2|\Psi_{1}|^{2}$ and $n_{\mathrm{s}2}=2|\Psi_{2}|^{2}$ accordingly), where these two condensates are coupled by both the internal proximity effect $\epsilon\left(\Psi_{1}^{+}\Psi_{2}+\Psi_{1}\Psi_{2}^{+}\right)$ and the ``drag'' effect $\eta\left(\nabla\Psi_{1}\nabla\Psi_{2}^{+}+\nabla\Psi_{1}^{+}\nabla\Psi_{2}\right)$~\cite{grig1,asker7,asker2,yerin1}. If we switch off the interband interactions $\epsilon=0$ and $\eta=0$, then we will have two independent superconductors with different critical temperatures $T_{c1}$ and $T_{c2}$ because the intraband interactions can be different. The sign of $\epsilon$ determines the equilibrium phase difference of the OP $|\Psi_{1}|\re^{\ri\theta_{1}}$ and $|\Psi_{2}|\re^{\ri\theta_{2}}$: $\theta_{1}-\theta_{2}=0$, if $\epsilon<0$, $|\theta_{1}-\theta_{2}|=\piup$, if $\epsilon>0$. The case $\epsilon<0$ corresponds to attractive interband interaction (for example, in $\mathrm{MgB}_{2}$), the case $\epsilon>0$ corresponds to repulsive interband interaction (for example, in iron-based superconductors). It should be noted that the effect of interband coupling $\epsilon\neq 0$, even if the coupling is weak, is nonperturbative: the application of a weak interband coupling washes out all OP up to a new critical temperature~\cite{grig3}.

In the work~\cite{grig3} there were investigated normal oscillations of internal degrees of freedom (Higgs mode and Goldstone mode) of two-band superconductors using generalization of the extended time-dependent Ginzburg-Landau (ETDGL) theory~\cite{grig2}, for the case of two coupled OP by both the internal proximity effect and the drag effect. It is demonstrated that, due to the internal proximity effect, the Goldstone mode splits into two branches: common mode oscillations with acoustic spectrum, which is absorbed by the gauge field, and anti-phase oscillations with an energy gap (mass) in the spectrum determined with the interband coupling $\epsilon$, which can be associated with the Leggett mode. Analogously, due to the internal proximity effect, Higgs oscillations also split into two branches. The energy gap of the common mode vanishes at a critical temperature $T_{c}$. For another anti-phase mode, its energy gap does not vanish at $T_{c}$ and is determined by the interband coupling $\epsilon$. It is demonstrated that the second branch of Higgs mode is nonphysical [since $|\Delta_{1,2}(T_{c})|=0$, then the mass of Higgs mode must be $m_{\rm H}(T_{c})=0$], and it, together with the Leggett mode, can be removed by special choice of the coefficient at the ``drag'' term in Lagrangian: $\eta^{2}={1}/{m_{1}m_{2}}$, $\eta\epsilon<0$ (where $m_{1,2}$ are electron masses in each band). Such a choice permits only one coherence length, thereby prohibiting the so-called type-1.5 superconductors. Thus, the drag effect is principally important: by special choice of the coefficient $\eta$ we ensure correct properties of the collective excitations in two-band superconductors. Experimental data of references~\cite{pono1,pono2} on the effect of resonant enhancement of the current through a Josephson junction between two-band superconductors is analyzed. It is demonstrated that the data can be explained by the coupling of Josephson oscillations  with Higgs oscillations of two-band superconductors $\hbar\omega=\sqrt{|\Delta_{1}||\Delta_{2}|}\propto\sqrt{|\Psi_{1}||\Psi_{2}|}$, and hence, \emph{these experiments cannot be considered as experimental confirmation of the Leggett mode}.

The physics of three-band SC systems (for example, some ferropnictides LiFeAs, NaFeAs,  Ba$_{1-x}$K$_{x}$Fe$_{2}$As$_{2}$~\cite{kuzm1,kuzm2,kord} and strontium ruthenate $\mathrm{Sr_{2}RuO_{4}}$~\cite{scaff}) is much richer and more complicated than the physics of two-band superconductors. In the three-band case, the equilibrium phase differences are not only $0$ or $\piup$, but they can be non-integer numbers of $\piup$ depending on the signs of the interband interactions $\epsilon_{ik}$~\cite{stanev1,stanev2,tanaka}. Thus, the equilibrium values of OP $\Psi_{1,2}$ in two-band superconductors are assumed to be real in the absence of current and magnetic field, although for three-band superconductors it is not always possible to make all OP $\Psi_{1,2,3}$ real, for example, when all interband couplings are repulsive ($\epsilon_{12}>0$, $\epsilon_{13}>0$, $\epsilon_{23}>0$) or when one coupling is repulsive but the other two are attractive (for example, $\epsilon_{12}>0$, $\epsilon_{13}<0$, $\epsilon_{23}<0$). As a consequence, the chiral ground state, frustration and the time-reversal symmetry breaking (TRSB)~\cite{stanev1,tanaka,stanev2,stanev3,babaev2,maiti,wilson,dias,yerin3}, the massless Leggett mode~\cite{lin1,kobay}, topological excitations~\cite{lin2,yerin4,babaev3}, type-1.5 regimes~\cite{babaev1} can occur. In addition, the TRSB state and the frustration essentially effects the Josephson current and magnetic penetration depth in junctions between three-band superconductors~\cite{hu,huang} and between single-band and three-band superconductors~\cite{yerin2,asker3,yerin5}. As demonstrated in references~\cite{stanev2,babaev1}, unlike two-band superconductors, the Higgs modes and the Leggett modes can be hybridized. Furthermore, the effect of hybridization is essential if the interband coupling is strong. The dynamics of two- and three-band systems was generalized for multi-band systems in reference~\cite{hase}, where it was demonstrated that there are 2 massive Leggett modes and $N-3$ massless Leggett modes in $N$-band systems ($N>2$). At the same time, in most works on multi-band superconductivity, the drag effect has not been taken into account, and, respectively, its role remains unknown.

Proceeding from the aforesaid, we aim at obtaining the spectrum of normal oscillations of internal degrees of freedom (Higgs modes and Goldstone modes), the coherence length, the ``light'' speed $\upsilon$ and the magnetic penetration depth using the method developed in the work~\cite{grig3} for the case of three order parameters coupled by both the internal proximity effect and the drag effect.



\section{Stationary regime}\label{stat}

Three-band superconductors are characterized by three OP-``wave functions'' $\Psi_{1}=|\Psi_{1}|\re^{\ri\theta_{1}}$, $\Psi_{2}=|\Psi_{2}|\re^{\ri\theta_{2}}$, $\Psi_{3}=|\Psi_{3}|\re^{\ri\theta_{3}}$ corresponding to the  condensates of Cooper pairs in each band, so that the densities of SC electrons are $n_{\mathrm{s}1}=2|\Psi_{1}|^{2}$, $n_{\mathrm{s}2}=2|\Psi_{2}|^{2}$, $n_{\mathrm{s}3}=2|\Psi_{3}|^{2}$, accordingly. In a bulk isotropic three-band superconductor, the Ginzburg-Landau free energy functional can be written as:
\begin{eqnarray}\label{1.1}
    F&=&\int \rd^{3}r\left[\frac{\hbar^{2}}{4m_{1}}\left|D\Psi_{1}\right|^{2}+\frac{\hbar^{2}}{4m_{2}}\left|D\Psi_{2}\right|^{2}
    +\frac{\hbar^{2}}{4m_{3}}\left|D\Psi_{3}\right|^{2}
    +\frac{\hbar^{2}}{4}\eta_{12}\left\{D\Psi_{1}(D\Psi_{2})^{+}+(D\Psi_{1})^{+}D\Psi_{2}\right\}\right.\nonumber\\
    &+&\frac{\hbar^{2}}{4}\eta_{13}\left\{D\Psi_{1}(D\Psi_{3})^{+}+(D\Psi_{1})^{+}D\Psi_{3}\right\}
    +\frac{\hbar^{2}}{4}\eta_{23}\left\{D\Psi_{2}(D\Psi_{3})^{+}+(D\Psi_{2})^{+}D\Psi_{3}\right\}\nonumber\\
    &+&a_{1}\left|\Psi_{1}\right|^{2}+a_{2}\left|\Psi_{2}\right|^{2}+a_{3}\left|\Psi_{3}\right|^{2}
    +\frac{b_{1}}{2}\left|\Psi_{1}\right|^{4}+\frac{b_{2}}{2}\left|\Psi_{2}\right|^{4}+\frac{b_{3}}{2}\left|\Psi_{3}\right|^{4}
    \nonumber\\
    &+&\left.\epsilon_{12}\left(\Psi_{1}^{+}\Psi_{2}+\Psi_{1}\Psi_{2}^{+}\right)+\epsilon_{13}\left(\Psi_{1}^{+}\Psi_{3}+\Psi_{1}\Psi_{3}^{+}\right)
    +\epsilon_{23}\left(\Psi_{2}^{+}\Psi_{3}+\Psi_{2}\Psi_{3}^{+}\right)+\frac{\mathbf{H}^{2}}{8\piup}\right],
\end{eqnarray}
where $D\equiv\nabla-\ri({2e}/{c\hbar})\mathbf{A}$ is a covariant gradient operator, ${\mathbf{H}^{2}}/{8\piup}={(\mathrm{curl}\,\mathbf{A})^{2}}/{8\piup}$ is the energy of magnetic field, $m_{1,2,3}$ denotes the effective mass of carriers in the corresponding band, the coefficients $a_{1,2,3}$ are given as $a_{i}=\gamma_{i}(T-T_{ci})$, where $\gamma_{i}$ are some constants, the coefficients $b_{1,2,3}$ are independent of temperature, the coefficients $\epsilon_{ij}$ and $\eta_{ij}$ describe the interband coupling of the OP (proximity effect) and their gradients (drag effect), respectively. If we switch off the interband interactions $\epsilon_{1,2,3}=0$ and $\eta_{1,2,3}=0$, then we will have three independent superconductors with different critical temperatures $T_{c1}$, $T_{c2}$, $T_{c3}$.

The potential  $V_{0}=\sum_{i=1}^{3}a_{i}\left|\Psi_{i}\right|^{2}+\frac{b_{i}}{2}\left|\Psi_{i}\right|^{4}$ is a sum of independent potentials of each condensate.  This energy is invariant under any phase rotation. Since the condensates in three-band superconductors are coupled by the Josephson terms $\epsilon_{ik}\left(\Psi_{i}^{+}\Psi_{k}+\Psi_{i}\Psi_{k}^{+}\right)=\epsilon_{ik}|\Psi_{i}||\Psi_{k}|\cos(\theta_{i}-\theta_{k})$, the broken $U(1)$ symmetry of the ground state in each band is shared throughout the system: the presence of the condensate $\langle\Psi_{i}\rangle\neq 0$ in some band induces the Cooper condensation in other bands $\langle\Psi_{k}\rangle\neq 0$,  that is the internal proximity effect takes place. At the same time, the Josephson terms breaks the global $U(1)$ gauge invariance, because these terms depend on the phase differences $\theta_{i}-\theta_{k}$, that is the Josephson terms have a physical sense as the interference between the condensates $\Psi_{1}$, $\Psi_{2}$, $\Psi_{3}$. Hence, the phase difference modes (the Leggett modes) acquire masses because the phase differences are fixed near the minima of the Josephson potential.

It should be noted that the free energy (\ref{1.1}) does not have the symmetry $U(1)\times Z_{2}$, unlike the statement in references~\cite{babaev1,babaev2}. Indeed, the transformation $Z_{2}$ makes the sign change of any two condensates, for example, $\Psi_{1}\rightarrow-\Psi_{1}$,  $\Psi_{2}\rightarrow-\Psi_{2}$ (which corresponds to the phase change $\theta_{1}\rightarrow\theta_{1}\pm\piup$,  $\theta_{2}\rightarrow\theta_{2}\pm\piup$), but the third condensate does not change its sign $\Psi_{3}\rightarrow\Psi_{3}$ (corresponding transformations are illustrated in appendix~\ref{symm}). Obviously, the sum of the Josephson terms in the free energy (\ref{1.1}) are not invariant under this transformation. It should be noted that the considered model can be referred to as the three-Higgs-doublet models (3HDM)~\cite{keus}, but without any specific symmetry in the above sense. For clarity, we present some invariant potentials under the simplest transformations in appendix~\ref{symm}. Thus, the potential~$V_{0}$ has $U(1)\times U(1)$ global gauge symmetry, but it is fully broken by the Josephson terms. In reference~\cite{hase}, the total rule was formulated: in the $N$-band system, the global symmetry $U(1)^{N-1}$ is broken by the Josephson terms to $U(1)^{N-3}$ symmetry. Thus, in $N>3$-band system, $N-3$ massless Leggett modes must be present. Ultimately, the system described with the free energy~(\ref{1.1}) is invariant under common $U(1)$ gauge transformation only, i.e., when each OP is turned by the same phase~$\theta$:  $\Psi_{k}\rightarrow\Psi_{k}\re^{\ri\theta}$. Hence, as demonstrated for two-band superconductors in reference~\cite{grig3}, the common mode phase oscillations are absorbed by the gauge field, although oscillations of the phase differences $\theta_{i}-\theta_{k}$ occur. The role of the intergradient couplings is the same as the Josephson coupling and does not introduce anything new at this stage.

\begin{figure}[h]
	\centerline{\includegraphics[width=0.45\textwidth]{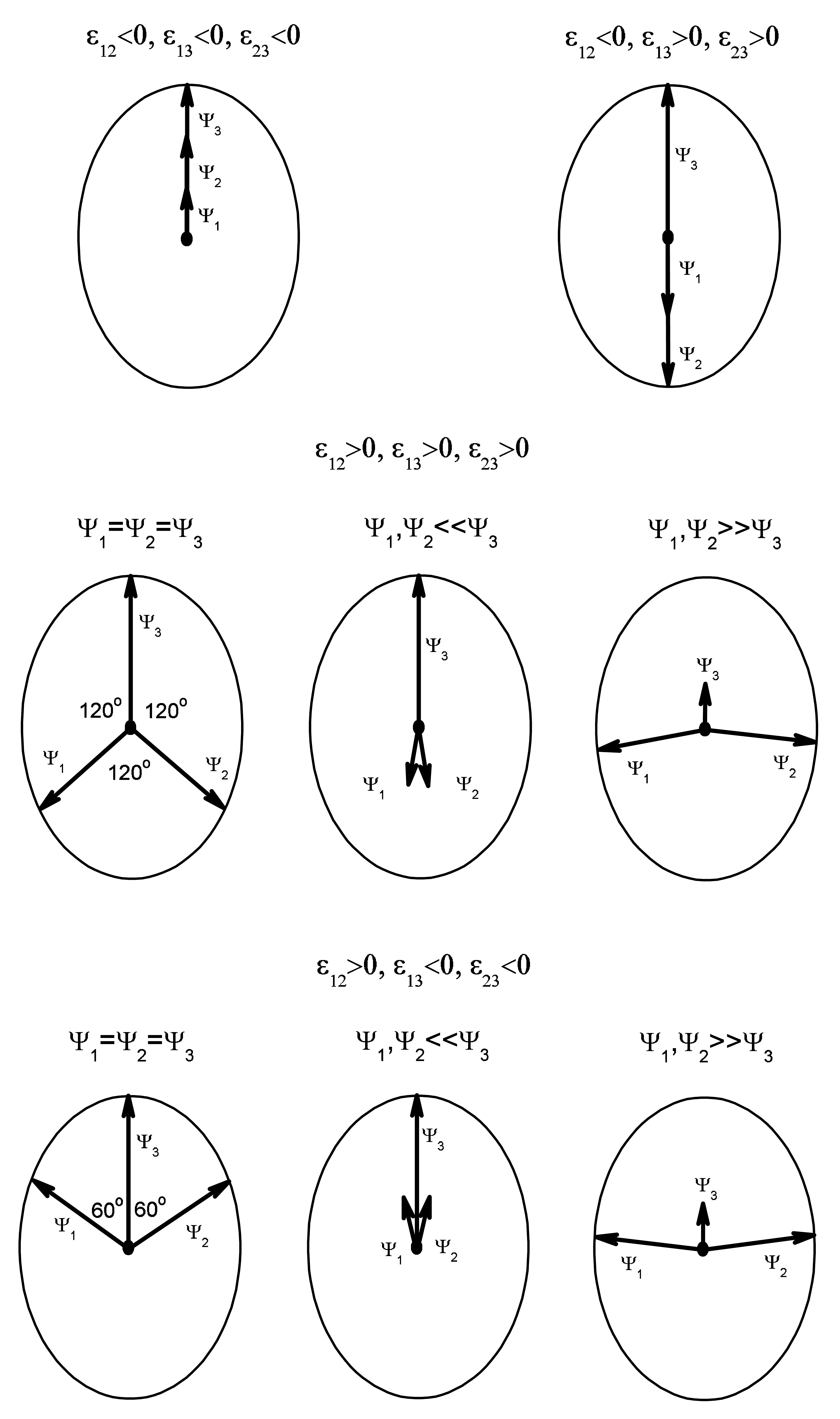}}
	\caption{The configurations of mutual arrangement of the OP $\Psi_{1}$, $\Psi_{2}$, $\Psi_{3}$ corresponding to some limit cases. It is supposed that the weak interband coupling $|\epsilon_{12}|=|\epsilon_{13}|=|\epsilon_{23}|\ll|a_{i}(0)|$.}
	\label{Fig1}
\end{figure}

Minimization of the free energy functional with respect to the OP, if $\nabla \Psi_{1,2,3}=0$ and $\mathbf{A}=0$, gives
\begin{eqnarray}\label{1.2}
  a_{1}\Psi_{1}+\epsilon_{12}\Psi_{2}+\epsilon_{13}\Psi_{3}+b_{1}|\Psi_{1}|^{2}\Psi_{1}=0,\nonumber \\
  a_{2}\Psi_{2}+\epsilon_{12}\Psi_{1}+\epsilon_{23}\Psi_{3}+b_{2}|\Psi_{2}|^{2}\Psi_{2}=0,\nonumber \\
  a_{3}\Psi_{3}+\epsilon_{13}\Psi_{1}+\epsilon_{23}\Psi_{2}+b_{3}|\Psi_{3}|^{2}\Psi_{3}=0. 
\end{eqnarray}
Equation~(\ref{1.2}) can be rewritten in a form:
\begin{eqnarray}\label{1.2a}
  a_{1}|\Psi_{1}|+\epsilon_{12}|\Psi_{2}|\re^{\ri(\theta_{2}-\theta_{1})}+\epsilon_{13}|\Psi_{3}|\re^{\ri(\theta_{3}-\theta_{1})}+b_{1}|\Psi_{1}|^{3}=0, \nonumber\\
  a_{2}|\Psi_{2}|+\epsilon_{12}|\Psi_{1}|\re^{\ri(\theta_{1}-\theta_{2})}+\epsilon_{23}|\Psi_{3}|\re^{\ri(\theta_{3}-\theta_{2})}+b_{1}|\Psi_{2}|^{3}=0, \nonumber\\
  a_{3}|\Psi_{3}|+\epsilon_{13}|\Psi_{1}|\re^{\ri(\theta_{1}-\theta_{3})}+\epsilon_{23}|\Psi_{2}|\re^{\ri(\theta_{2}-\theta_{3})}+b_{1}|\Psi_{3}|^{3}=0,
\end{eqnarray}
or in an expanded form:
\begin{eqnarray}\label{1.2b}
  &&a_{1}|\Psi_{1}|+\epsilon_{12}|\Psi_{2}|\cos(\theta_{2}-\theta_{1})+\epsilon_{13}|\Psi_{3}|\cos(\theta_{3}-\theta_{1})+b_{1}|\Psi_{1}|^{3}=0, \nonumber\\
  &&a_{2}|\Psi_{2}|+\epsilon_{12}|\Psi_{1}|\cos(\theta_{1}-\theta_{2})+\epsilon_{23}|\Psi_{3}|\cos(\theta_{3}-\theta_{2})+b_{1}|\Psi_{2}|^{3}=0, \nonumber\\
  &&a_{3}|\Psi_{3}|+\epsilon_{13}|\Psi_{1}|\cos(\theta_{1}-\theta_{3})+\epsilon_{23}|\Psi_{2}|\cos(\theta_{2}-\theta_{3})+b_{1}|\Psi_{3}|^{3}=0, \nonumber\\
  &&\epsilon_{12}|\Psi_{2}|\sin(\theta_{2}-\theta_{1})+\epsilon_{13}|\Psi_{3}|\sin(\theta_{3}-\theta_{1})=0, \nonumber\\
  &&\epsilon_{12}|\Psi_{1}|\sin(\theta_{1}-\theta_{2})+\epsilon_{23}|\Psi_{3}|\sin(\theta_{3}-\theta_{2})=0, \nonumber\\
  &&\epsilon_{13}|\Psi_{1}|\sin(\theta_{1}-\theta_{3})+\epsilon_{23}|\Psi_{2}|\sin(\theta_{2}-\theta_{3})=0.
\end{eqnarray}
Hence, possible signs of $\epsilon_{ik}$ and the phase differences of the OP $|\Psi_{1}|\re^{\ri\theta_{1}}$,  $|\Psi_{2}|\re^{\ri\theta_{2}}$,  $|\Psi_{3}|\re^{\ri\theta_{3}}$ are:
\begin{eqnarray}\label{1.4a}
      \epsilon_{ik}<0,\quad\epsilon_{il}<0,\quad\epsilon_{kl}<0, & \mathrm{then}&\cos(\theta_{i}-\theta_{k})=\cos(\theta_{i}-\theta_{l})=\cos(\theta_{k}-\theta_{l})=1,  \nonumber\\
      \epsilon_{ik}<0,\quad\epsilon_{il}>0,\quad\epsilon_{kl}>0, & \mathrm{then}&\cos(\theta_{i}-\theta_{k})=1,\quad\cos(\theta_{i}-\theta_{l})=\cos(\theta_{k}-\theta_{l})=-1, \qquad
\end{eqnarray}
for the cases, when $\epsilon_{12}\epsilon_{13}\epsilon_{23}<0$. Therefore, OP $\Psi_{1}$, $\Psi_{2}$, $\Psi_{3}$ can be assumed to be real simultaneously, as in two-band superconductors. The cases $\epsilon_{12}\epsilon_{13}\epsilon_{23}>0$ need numerical solution of equation~(\ref{1.2b}). As a result, the phase differences $\theta_{ik}\equiv\theta_{i}-\theta_{k}$ can be functions of temperature $\theta_{ik}(T)$. Only in the case of absolutely symmetrical bands $a_{1}=a_{2}=a_{3}$,  $b_{1}=b_{2}=b_{3}$,  $|\epsilon_{12}|=|\epsilon_{13}|=|\epsilon_{23}|$  we obtain
\begin{eqnarray}\label{1.4b}
    &&\epsilon_{ik}>0,\quad\epsilon_{il}>0,\quad\epsilon_{kl}>0, \,\,\, \mathrm{then}\,\,\, \cos(\theta_{i}-\theta_{k})=\cos(\theta_{i}-\theta_{l})=\cos(\theta_{k}-\theta_{l})=-{1}/{2},  \nonumber\\
   &&\epsilon_{ik}>0,\,\,\, \epsilon_{il}<0,\,\,\, \epsilon_{kl}<0,\,\,\, \mathrm{then}\,\,\, \cos(\theta_{i}-\theta_{k})=-{1}/{2},\,\,\, \cos(\theta_{i}-\theta_{l})=\cos(\theta_{k}-\theta_{l})={1}/{2}.\qquad 
\end{eqnarray}
As an approximation in the case of weak coupling $|\epsilon_{12}|,|\epsilon_{13}|,|\epsilon_{23}|\ll |a_{1}(0)|,|a_{2}(0)|,|a_{3}(0)|$, we can assume $|\Psi_{i}(0)|=\sqrt{{|a_{i}(0)|}/{b_{i}}}$ and then substitute them in equation~(\ref{1.2b}) to find the angles $\theta_{i}-\theta_{k}$. Then, using the found phase differences, we find new $|\Psi_{i}(T)|$ from equation~(\ref{1.2b}) for all temperatures. Possible configurations corresponding to some limit cases are illustrated in figure~\ref{Fig1}. However, it should be noted that, as demonstrated in reference~\cite{yerin3} by numerical calculations, in the case $\epsilon_{12}\epsilon_{13}\epsilon_{23}>0$, the regime of nontrivial phase differences $\theta_{2}-\theta_{1},\theta_{3}-\theta_{1}\neq 0,\piup$ (that is the TRSB state can be realized) exists only within a relative small volume in the six-dimensional parameter space $(|\Psi_{i}|,\epsilon_{ik})$.

\begin{figure}[h]
	\centerline{\includegraphics[width=0.5\textwidth]{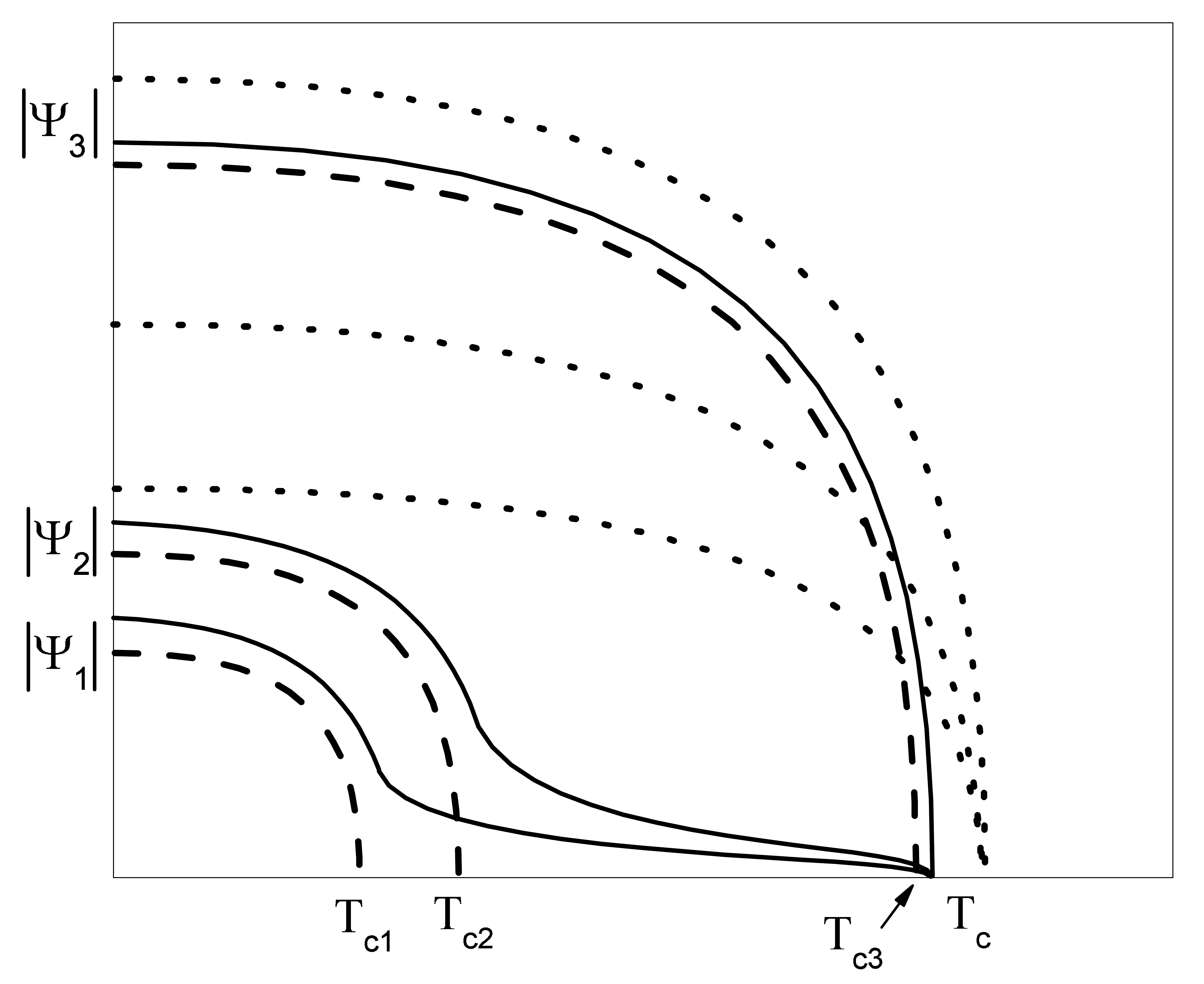}}
	\caption{The sketch of temperature dependencies of OP $\Psi_{1}(T)$, $\Psi_{2}(T)$, $\Psi_{3}(T)$ as solutions of equation~(\ref{1.2b}), if the interband couplings are absent, i.e., $\epsilon_{ik}=0$ (dash lines), and if the weak interband interaction takes place, i.e., $\epsilon_{ik}\neq 0$, $|\epsilon_{ik}|\ll |a_{1}(0)|$ (solid lines). The application of the weak interband coupling washes out the smaller parameters $\Psi_{1,2}$ up to a new critical temperature $T_{c}\gg T_{c1}$. The effect on the larger parameter $\Psi_{3}$ is not so essential. As the couplings $|\epsilon_{ik}|$ increase, $\Psi_{1,2,3}(T)$ take the forms shown with dot lines.}
	\label{Fig2}
\end{figure}

Near critical temperature $T_{c}$ we have $|\Psi_{1,2,3}|^{2}\rightarrow 0$. Hence, we can find the critical temperature equating to zero the determinant of the linearized system (\ref{1.2}):
\begin{equation}\label{1.3a}
a_{1}a_{2}a_{3}-a_{1}\epsilon_{23}^{2}-a_{2}\epsilon_{13}^{2}-a_{3}\epsilon_{12}^{2}+2\epsilon_{12}\epsilon_{13}\epsilon_{23}=0.
\end{equation}
In an equivalent way, we can find the critical temperature equating to zero the determinant of the linearized system of the first three equations from equation~(\ref{1.2b}):
\begin{eqnarray}\label{1.3}
a_{1}a_{2}a_{3}&-&a_{1}\epsilon_{23}^{2}\cos^{2}(\theta_{1}-\theta_{2})
-a_{2}\epsilon_{13}^{2}\cos^{2}(\theta_{1}-\theta_{3})-a_{3}\epsilon_{12}^{2}\cos^{2}(\theta_{1}-\theta_{2})\nonumber\\
&+&2\epsilon_{12}\epsilon_{13}\epsilon_{23}\cos(\theta_{1}-\theta_{2})\cos(\theta_{1}-\theta_{3})\cos(\theta_{2}-\theta_{3})=0,
\end{eqnarray}
where phase differences $\theta_{i}-\theta_{k}$ are \emph{equilibrium} values, that is, those that ensure the coincidence of the solutions of the equations (\ref{1.3a}) and (\ref{1.3}), and the critical temperature $T_{c}$ of the system is the largest of these solutions. 
Solving any of these equations we find $T_{c}>T_{c1},T_{c2},T_{c3}$, and besides $T_{c}\left(\epsilon_{12}\epsilon_{13}\epsilon_{23}<0\right)>T_{c}\left(\epsilon_{12}\epsilon_{13}\epsilon_{23}>0\right)$. The case $\epsilon_{ik}<0$ corresponds to attractive interband interaction, the case $\epsilon_{ik}>0$ corresponds to repulsive interband interaction. For symmetrical bands $T_{c1}=T_{c2}=T_{c3}\equiv T_{c123}$,  $\gamma_{1}=\gamma_{2}=\gamma_{3}\equiv\gamma$ and the same modulus of interband interactions $|\epsilon_{12}|=|\epsilon_{13}|=|\epsilon_{23}|\equiv\epsilon>0$ equation~(\ref{1.3}) is reduced to
\begin{eqnarray}\label{1.4}
\epsilon_{12}\epsilon_{13}\epsilon_{23}<0 &\Rightarrow& (a+\epsilon)^{2}(a-2\epsilon)=0 \,\,\,\,\Rightarrow\,\,\,\, T_{c}=T_{c123}+{2\epsilon}/{\gamma},\nonumber\\
\epsilon_{12}\epsilon_{13}\epsilon_{23}>0 &\Rightarrow&\left(a+\frac{\epsilon}{2}\right)^{2}(a-\epsilon)=0  \,\,\,\,\Rightarrow\,\,\,\, T_{c}=T_{c123}+{\epsilon}/{\gamma}.
\end{eqnarray}
The solutions of equation~(\ref{1.2b}) are illustrated in figure~\ref{Fig2} for the case of strongly asymmetrical bands $T_{c1,c2}\ll T_{c3}$. As in the two-band system, the effect of interband coupling $\epsilon_{ik}\neq 0$, even if the coupling is weak $|\epsilon_{ik}|\ll |a_{1}(0)|$, is non-perturbative for the smaller OP $\Psi_{1,2}$ --- the application of the weak interband coupling washes out the smaller OP up to a new critical temperature $T_{c}\gg T_{c1,c2}$. At the same time, the effect on the largest parameter $\Psi_{3}$ is not so significant --- the application of the interband coupling slightly increases the critical temperature $T_{c}\gtrsim T_{c3}$ only.

Let us consider a superconductor in the weak magnetic field $\mathbf{A}(\mathbf{r})$ (i.e., $|\Psi|=\mathrm{const}$). Then, the free energy functional (\ref{1.1}) can be reduced to the form:
\begin{eqnarray}\label{1.6}
    F&=&\int \rd^{3}r \mathfrak{F}\equiv\int \rd^{3}r \left[\frac{\hbar^{2}}{4m_{1}}|\Psi_{1}|^{2}\left(\nabla\theta_{1}-\frac{2e}{\hbar c}\mathbf{A}\right)^{2}
    +\frac{\hbar^{2}}{4m_{2}}|\Psi_{2}|^{2}\left(\nabla\theta_{2}-\frac{2e}{\hbar c}\mathbf{A}\right)^{2}\right. \nonumber\\
    &+&\frac{\hbar^{2}}{4m_{3}}|\Psi_{3}|^{2}\left(\nabla\theta_{3}-\frac{2e}{\hbar c}\mathbf{A}\right)^{2}
    +\frac{\hbar^{2}}{2}\eta_{12}|\Psi_{1}||\Psi_{2}|
    \left(\nabla\theta_{1}-\frac{2e}{\hbar c}\mathbf{A}\right)\left(\nabla\theta_{2}-\frac{2e}{\hbar c}\mathbf{A}\right)\cos(\theta_{1}-\theta_{2})\nonumber\\
    &+&\frac{\hbar^{2}}{2}\eta_{13}|\Psi_{1}||\Psi_{3}|
    \left(\nabla\theta_{1}-\frac{2e}{\hbar c}\mathbf{A}\right)\left(\nabla\theta_{3}-\frac{2e}{\hbar c}\mathbf{A}\right)\cos(\theta_{1}-\theta_{3})\nonumber\\
    &+&\frac{\hbar^{2}}{2}\eta_{23}|\Psi_{2}||\Psi_{3}|
    \left(\nabla\theta_{2}-\frac{2e}{\hbar c}\mathbf{A}\right)\left(\nabla\theta_{3}-\frac{2e}{\hbar c}\mathbf{A}\right)\cos(\theta_{2}-\theta_{3})+\frac{(\mathrm{curl}\,\mathbf{A})^{2}}{8\piup}\nonumber\\
    &+&\left.\sum_{i=1}^{3}\left(a_{i}|\Psi_{i}|^{2}+\frac{b_{i}}{2}|\Psi_{i}|^{4}\right)+\sum_{i\neq k}\epsilon_{ik}\left(\Psi_{i}^{+}\Psi_{k}+\Psi_{i}\Psi_{k}^{+}\right)\right].
\end{eqnarray}
Corresponding Lagrange equation
\begin{equation}\label{1.7a}
  \mathrm{curl}\,\frac{\partial\mathfrak{F}}{\partial(\mathrm{curl}\,\mathbf{A})}-\frac{\partial\mathfrak{F}}{\partial\mathbf{A}}=0
\end{equation}
gives the supercurrent:
\begin{eqnarray}\label{1.7}
  \mathbf{J}&=&\frac{\hbar e}{m_{1}}|\Psi_{1}|^{2}\left(\nabla\theta_{1}-\frac{2e}{\hbar c}\mathbf{A}\right)+
  \frac{\hbar e}{m_{2}}|\Psi_{2}|^{2}\left(\nabla\theta_{2}-\frac{2e}{\hbar c}\mathbf{A}\right)+
  \frac{\hbar e}{m_{3}}|\Psi_{3}|^{2}\left(\nabla\theta_{3}-\frac{2e}{\hbar c}\mathbf{A}\right)\nonumber\\
  &+&\hbar e\eta_{12}|\Psi_{1}||\Psi_{2}|
  \left(\nabla\theta_{1}+\nabla\theta_{2}-2\frac{2e}{\hbar c}\mathbf{A}\right)\cos(\theta_{1}-\theta_{2})\nonumber\\
  &+&\hbar e\eta_{13}|\Psi_{1}||\Psi_{3}|
  \left(\nabla\theta_{1}+\nabla\theta_{3}-2\frac{2e}{\hbar c}\mathbf{A}\right)\cos(\theta_{1}-\theta_{3})\nonumber\\
  &+&\hbar e\eta_{23}|\Psi_{2}||\Psi_{3}|
  \left(\nabla\theta_{2}+\nabla\theta_{3}-2\frac{2e}{\hbar c}\mathbf{A}\right)\cos(\theta_{2}-\theta_{3}),
\end{eqnarray}
that can be rewritten in the following form:
\begin{eqnarray}\label{1.8}
  \mathbf{J}=\frac{\hbar e}{\mathfrak{m}_{1}}|\Psi_{1}|^{2}\nabla\theta_{1}+
  \frac{\hbar e}{\mathfrak{m}_{2}}|\Psi_{2}|^{2}\nabla\theta_{2}+\frac{\hbar e}{\mathfrak{m}_{3}}|\Psi_{3}|^{2}\nabla\theta_{3}
  -\left(\frac{2e^{2}}{\mathfrak{m}_{1}c}|\Psi_{1}|^{2}
  +\frac{2e^{2}}{\mathfrak{m}_{2}c}|\Psi_{2}|^{2}+\frac{2e^{2}}{\mathfrak{m}_{3}c}|\Psi_{3}|^{2}\right)\mathbf{A},
\end{eqnarray}
where $\mathfrak{m}_{i}$ is the effective mass of an electron in a band $i$ due to the drag effect:

\begin{equation}\label{1.9}
  \frac{1}{\mathfrak{m}_{i}}=\frac{1}{m_{i}}\left[1+\eta_{ik}m_{i}\frac{|\Psi_{k}|}{|\Psi_{i}|}\cos(\theta_{i}-\theta_{k})
  +\eta_{il}m_{i}\frac{|\Psi_{l}|}{|\Psi_{i}|}\cos(\theta_{i}-\theta_{l})\right].
\end{equation}

The magnetic field can be gauge transformed as
\begin{equation}\label{1.10}
    \mathbf{A}=\mathbf{A}'+\frac{\hbar c}{2e}\left(\alpha\nabla\theta_{1}+\beta\nabla\theta_{2}+\gamma\nabla\theta_{3}\right),
\end{equation}
where
\begin{equation}\label{1.11}
\alpha=\frac{{|\Psi_{1}|^{2}}/{\mathfrak{m}_{1}}}{\frac{|\Psi_{1}|^{2}}{\mathfrak{m}_{1}}+\frac{|\Psi_{2}|^{2}}{\mathfrak{m}_{2}}
+\frac{|\Psi_{3}|^{2}}{\mathfrak{m}_{3}}},\quad
\beta=\frac{{|\Psi_{2}|^{2}}/{\mathfrak{m}_{2}}}{\frac{|\Psi_{1}|^{2}}{\mathfrak{m}_{1}}+\frac{|\Psi_{2}|^{2}}{\mathfrak{m}_{2}}
+\frac{|\Psi_{3}|^{2}}{\mathfrak{m}_{3}}},\quad
\gamma=\frac{{|\Psi_{3}|^{2}}/{\mathfrak{m}_{3}}}{\frac{|\Psi_{1}|^{2}}{\mathfrak{m}_{1}}+\frac{|\Psi_{2}|^{2}}{\mathfrak{m}_{2}}
+\frac{|\Psi_{3}|^{2}}{\mathfrak{m}_{3}}},
\end{equation}
so that
\begin{equation}\label{1.12}
    \alpha+\beta+\gamma=1,\quad\frac{|\Psi_{2}|^{2}}{\mathfrak{m}_{2}}\frac{|\Psi_{3}|^{2}}{\mathfrak{m}_{3}}\alpha
    =\frac{|\Psi_{1}|^{2}}{\mathfrak{m}_{1}}\frac{|\Psi_{3}|^{2}}{\mathfrak{m}_{3}}\beta
    =\frac{|\Psi_{1}|^{2}}{\mathfrak{m}_{1}}\frac{|\Psi_{2}|^{2}}{\mathfrak{m}_{2}}\gamma.
\end{equation}
Then, equation~(\ref{1.8}) is reduced to the London law:
\begin{eqnarray}\label{1.13}
  \mathbf{J}=-\left(\frac{2e^{2}}{\mathfrak{m}_{1}c}|\Psi_{1}|^{2}
  +\frac{2e^{2}}{\mathfrak{m}_{2}c}|\Psi_{2}|^{2}+\frac{2e^{2}}{\mathfrak{m}_{3}c}|\Psi_{3}|^{2}\right)\mathbf{A}\equiv
  -\frac{1}{\lambda^{2}}\mathbf{A}.
\end{eqnarray}
Thus, magnetic response of three-band superconductors is analogous to the response of single-band superconductors, but with contribution into SC density from each band $|\Psi_{i}|^{2}$ with the corresponding effective electron mass (\ref{1.9}), which is determined with the coefficients of the drag effect $\eta_{ik}$.

\section{Goldstone and Higgs oscillations in three-band superconductors}\label{Goldstone and Higgs}

\subsection{Ginzburg-Landau Lagrangian for three-band superconductors}

In general case, the OP $\Psi_{1,2,3}$ are both spatially inhomogeneous and they can change over time:
$\Psi_{1,2,3}=\Psi_{1,2,3}(\textbf{r},t)$. The OP in the modulus-phase representation are equivalent to two real fields each: modulus $\left|\Psi(\textbf{r},t)\right|$ and phase $\theta(\textbf{r},t)$:
\begin{equation}\label{2.1}
    \Psi_{1}(\textbf{r},t)=\left|\Psi_{1}(\textbf{r},t)\right|\re^{\ri\theta_{1}(\textbf{r},t)}, \quad\Psi_{2}(\textbf{r},t)=\left|\Psi_{2}(\textbf{r},t)\right|\re^{\ri\theta_{2}(\textbf{r},t)},
    \quad\Psi_{3}(\textbf{r},t)=\left|\Psi_{3}(\textbf{r},t)\right|\re^{\ri\theta_{3}(\textbf{r},t)}.
\end{equation}
For the stationary case $\Psi_{1,2,3}=\Psi_{1,2,3}(\textbf{r})$, the steady configuration of the field $\Psi_{1,2,3}(\textbf{r})$ minimizes the free energy functional (\ref{1.1}). For the nonstationary case $\Psi_{1,2,3}(\textbf{r},t)$, according to the method described in~\cite{grig2}, we consider some 4D Minkowski space $\{\upsilon t,\textbf{r}\}$, where the parameter $\upsilon$ plays the role of the ``light'' speed, which should be determined by the dynamical properties of the system. At the same time, the dynamics of conduction electrons remains non-relativistic. Then, the two-component scalar fields $\Psi_{1,2,3}(\textbf{r},t)$ minimize some action $S$ in the Minkowski space:
\begin{equation}\label{2.2}
    S=\frac{1}{\upsilon}\int\mathcal{L}\left(\Psi_{1},\Psi_{2},\Psi_{3},\Psi^{+}_{1},\Psi^{+}_{2},\Psi^{+}_{3}
    ,\widetilde{A}_{\mu},\widetilde{A}^{\mu}\right)\upsilon\, \rd t\, \rd^{3}r,
\end{equation}
where $\widetilde{A}_{\mu}=(\frac{c}{\upsilon}\varphi,-\mathbf{A})$, $\widetilde{A}^{\mu}=(\frac{c}{\upsilon}\varphi,\mathbf{A})$ are covariant and contravariant potential of electromagnetic field. The Lagrangian $\mathcal{L}$ is built by generalizing the density of free energy in equation~(\ref{1.1}) to the ``relativistic'' invariant form by substitution of covariant and contravariant differential operators:
\begin{equation}\label{2.3}
\widetilde{\partial}_{\mu}\equiv\left(\frac{1}{\upsilon}\frac{\partial}{\partial t},\nabla\right),
\quad\widetilde{\partial}^{\mu}\equiv\left(\frac{1}{\upsilon}\frac{\partial}{\partial t},-\nabla\right),
\end{equation}
instead of the gradient operators:
$\nabla\Psi\rightarrow\widetilde{\partial}_{\mu}\Psi$,  $\nabla\Psi^{+}\rightarrow\widetilde{\partial}^{\mu}\Psi^{+}$, and by substitution the covariant and contravariant operators in presence of electromagnetic field $A_{\mu}$:
\begin{equation}\label{2.3a}
D_{\mu}\equiv\widetilde{\partial}_{\mu}+\frac{\ri 2\widetilde{e}}{\hbar\upsilon}\widetilde{A}_{\mu},\quad D^{\mu}\equiv\widetilde{\partial}^{\mu}+\frac{\ri 2\widetilde{e}}{\hbar\upsilon}\widetilde{A}^{\mu},
\end{equation}
instead of the operators $D$ in the free energy functional (\ref{1.1}). Here, $\widetilde{e}=\frac{\upsilon}{c}e$, so that $\widetilde{e}\widetilde{A}_{\mu}=eA_{\mu}$. Then, the Lagrangian will be written as:
\begin{eqnarray}\label{2.4}
    \mathcal{L}&=&\frac{\hbar^{2}}{4m_{1}}D_{\mu}\Psi_{1}D^{\mu}\Psi_{1}^{+}+\frac{\hbar^{2}}{4m_{2}}D_{\mu}\Psi_{2}D^{\mu}\Psi_{2}^{+}
    +\frac{\hbar^{2}}{4m_{3}}D_{\mu}\Psi_{3}D^{\mu}\Psi_{3}^{+}\nonumber\\
    &+&\frac{\hbar^{2}}{4}\eta_{12}\left\{D_{\mu}\Psi_{1}(D^{\mu}\Psi_{2})^{+}+(D^{\mu}\Psi_{1})^{+}D_{\mu}\Psi_{2}\right\}\nonumber\\
    &+&\frac{\hbar^{2}}{4}\eta_{13}\left\{D_{\mu}\Psi_{1}(D^{\mu}\Psi_{3})^{+}+(D^{\mu}\Psi_{1})^{+}D_{\mu}\Psi_{3}\right\}
    +\frac{\hbar^{2}}{4}\eta_{23}\left\{D_{\mu}\Psi_{2}(D^{\mu}\Psi_{3})^{+}+(D^{\mu}\Psi_{2})^{+}D_{\mu}\Psi_{3}\right\}\nonumber\\
    &-&a_{1}\left|\Psi_{1}\right|^{2}-a_{2}\left|\Psi_{2}\right|^{2}-a_{3}\left|\Psi_{3}\right|^{2}
    -\frac{b_{1}}{2}\left|\Psi_{1}\right|^{4}-\frac{b_{2}}{2}\left|\Psi_{2}\right|^{4}-\frac{b_{3}}{2}\left|\Psi_{3}\right|^{4}
    \nonumber\\
    &-&\epsilon_{12}\left(\Psi_{1}^{+}\Psi_{2}+\Psi_{1}\Psi_{2}^{+}\right)-\epsilon_{13}\left(\Psi_{1}^{+}\Psi_{3}+\Psi_{1}\Psi_{3}^{+}\right)
    -\epsilon_{23}\left(\Psi_{2}^{+}\Psi_{3}+\Psi_{2}\Psi_{3}^{+}\right)-\frac{1}{16\piup}\widetilde{F}_{\mu\nu}\widetilde{F}^{\mu\nu},
\end{eqnarray}
where the same speed $\upsilon$ is used for the condensates $\Psi_{1,2,3}$ with the masses $m_{1,2,3}$, accordingly. The speed~$\upsilon$ plays the role of the speed of light in SC medium, and it will be found below. $\widetilde{F}_{\mu\nu}=\widetilde{\partial}_{\mu}\widetilde{A}_{\nu}-\widetilde{\partial}_{\nu}\widetilde{A}_{\mu}$ is the Faraday tensor.

The modulus-phase representation (\ref{2.1}) can be considered as the local gauge $U(1)$ transformation $\Psi_{i}\rightarrow|\Psi_{i}|$. Then, the gauge field $\widetilde{A}_{\mu}$ should be transformed as
\begin{equation}\label{2.6}
    \widetilde{A}_{\mu}'=\widetilde{A}_{\mu}+\frac{\hbar\upsilon}{2\widetilde{e}}\left(\alpha\widetilde{\partial}_{\mu}\theta_{1}
    +\beta\widetilde{\partial}_{\mu}\theta_{2}+\gamma\widetilde{\partial}_{\mu}\theta_{3}\right),
\end{equation}
where coefficients $\alpha$, $\beta$, $\gamma$ are determined with equation~(\ref{1.11}). The transformation (\ref{2.6}) excludes the phases $\theta_{1}$, $\theta_ {2}$, $\theta_ {3}$ [using properties (\ref{1.12})] from Lagrangian (\ref{2.4}) individually leaving only their differences:

\begin{eqnarray}\label{2.7}
    \mathcal{L}&=&\frac{\hbar^{2}}{4m_{1}}D_{\mu}|\Psi_{1}|D^{\mu}|\Psi_{1}|+\frac{\hbar^{2}}{4m_{2}}D_{\mu}|\Psi_{1}|D^{\mu}|\Psi_{2}|
    +\frac{\hbar^{2}}{4m_{3}}D_{\mu}|\Psi_{3}|D^{\mu}|\Psi_{3}| \nonumber\\
    &+&\frac{\hbar^{2}}{2}\eta_{12}D_{\mu}|\Psi_{1}|D^{\mu}|\Psi_{2}|\cos(\theta_{1}-\theta_{2})
    +\frac{\hbar^{2}}{2}\eta_{13}D_{\mu}|\Psi_{1}|D^{\mu}|\Psi_{3}|\cos(\theta_{1}-\theta_{3}) \nonumber\\
    &+&\frac{\hbar^{2}}{2}\eta_{23}D_{\mu}|\Psi_{2}|D^{\mu}|\Psi_{3}|\cos(\theta_{2}-\theta_{3})-
    2\epsilon_{12}|\Psi_{1}||\Psi_{2}|\cos(\theta_{1}-\theta_{2})
    -2\epsilon_{13}|\Psi_{1}||\Psi_{3}|\cos(\theta_{1}-\theta_{3})\nonumber\\
    &-&2\epsilon_{23}|\Psi_{2}||\Psi_{3}|\cos(\theta_{2}-\theta_{3})
    +\frac{\hbar^{2}}{4}\left[\frac{|\Psi_{1}|^{2}}{m_{1}}\beta^{2}+\frac{|\Psi_{2}|^{2}}{m_{2}}\alpha^{2}
    +2\eta_{12}|\Psi_{1}||\Psi_{2}|\alpha\beta\cos(\theta_{1}-\theta_{2})\right]   \nonumber\\
    &\times&\widetilde{\partial}_{\mu}\left(\theta_{1}-\theta_{2}\right)\widetilde{\partial}^{\mu}\left(\theta_{1}-\theta_{2}\right)
    +\frac{\hbar^{2}}{4}\left[\frac{|\Psi_{1}|^{2}}{m_{1}}\gamma^{2}+\frac{|\Psi_{3}|^{2}}{m_{3}}\alpha^{2}
    +2\eta_{13}|\Psi_{1}||\Psi_{3}|\alpha\gamma\cos(\theta_{1}-\theta_{3})\right] \nonumber\\
    &\times& \widetilde{\partial}_{\mu}\left(\theta_{1}-\theta_{3}\right)\widetilde{\partial}^{\mu}\left(\theta_{1}-\theta_{3}\right)
    +\frac{\hbar^{2}}{4}\left[\frac{|\Psi_{2}|^{2}}{m_{2}}\gamma^{2}+\frac{|\Psi_{3}|^{2}}{m_{3}}\beta^{2}
    +2\eta_{23}|\Psi_{2}||\Psi_{3}|\beta\gamma\cos(\theta_{2}-\theta_{3})\right]
    \widetilde{\partial}_{\mu}\left(\theta_{2}-\theta_{3}\right) \nonumber\\ &\times&\widetilde{\partial}^{\mu}\left(\theta_{2}-\theta_{3}\right)
    -\frac{\hbar^{2}}{4}\bigg[\frac{|\Psi_{1}|^{2}}{m_{1}}2\gamma\beta
    -2\eta_{12}|\Psi_{1}||\Psi_{2}|\alpha\gamma\cos(\theta_{1}-\theta_{2})-2\eta_{13}|\Psi_{1}||\Psi_{3}|\alpha\beta\cos(\theta_{1}-\theta_{3}) \nonumber\\
    &+&2\eta_{23}|\Psi_{2}||\Psi_{3}|\alpha^{2}\cos(\theta_{2}-\theta_{3})\bigg]
    \widetilde{\partial}_{\mu}\left(\theta_{1}-\theta_{2}\right)\widetilde{\partial}^{\mu}\left(\theta_{1}-\theta_{3}\right)
    -\frac{\hbar^{2}}{4}\bigg[\frac{|\Psi_{2}|^{2}}{m_{2}}2\alpha\gamma \nonumber\\
    &+&2\eta_{12}|\Psi_{1}||\Psi_{2}|\beta\gamma\cos(\theta_{1}-\theta_{2})-2\eta_{13}|\Psi_{1}||\Psi_{3}|\beta^{2}\cos(\theta_{1}-\theta_{3})
    +2\eta_{23}|\Psi_{2}||\Psi_{3}|\alpha\beta\cos(\theta_{2}-\theta_{3})\bigg]\nonumber\\
    &\times&\widetilde{\partial}_{\mu}\left(\theta_{1}-\theta_{2}\right)\widetilde{\partial}^{\mu}\left(\theta_{2}-\theta_{3}\right)
    -\frac{\hbar^{2}}{4}\bigg[\frac{|\Psi_{3}|^{2}}{m_{3}}2\alpha\beta +2\eta_{12}|\Psi_{1}||\Psi_{2}|\gamma^{2}\cos(\theta_{1}-\theta_{2}) \nonumber\\
    &+&2\eta_{13}|\Psi_{1}||\Psi_{3}|\beta\gamma\cos(\theta_{1}-\theta_{3})
    -2\eta_{23}|\Psi_{2}||\Psi_{3}|\alpha\gamma\cos(\theta_{2}-\theta_{3})\bigg]
    \widetilde{\partial}_{\mu}\left(\theta_{1}-\theta_{3}\right)\widetilde{\partial}^{\mu}\left(\theta_{2}-\theta_{3}\right) \nonumber\\
    &+&\mathcal{L}\left(|\Psi_{1}|,|\Psi_{2}|,|\Psi_{3}|,\widetilde{F}_{\mu\nu}\widetilde{F}^{\mu\nu}\right).
\end{eqnarray}
Here, $\mathcal{L}\left(|\Psi_{1}|,|\Psi_{2}|,|\Psi_{3}|,\widetilde{F}_{\mu\nu}\widetilde{F}^{\mu\nu}\right)\equiv -\sum_{i=1}^{3}\left(a|\Psi_{i}|^{2}+\frac{b}{2}|\Psi_{i}|^{4}\right)-\frac{1}{16\piup}\widetilde{F}_{\mu\nu}\widetilde{F}^{\mu\nu}$ is the sum of terms of the Lagrangian, which do not depend on the phases $\theta_{i}$: single-band potential energies and Lagrangian of electromagnetic field.
Thus, the gauge field $\widetilde{A}_{\mu}$ absorbs the Goldstone bosons $\theta_{1,2,3}$ so that the Lagrangian becomes dependent on the phase differences $\theta_{1}-\theta_{2}$, $\theta_{1}-\theta_{3}$, $\theta_{2}-\theta_{3}$ only. At the same time, the phase differences are not normal coordinates, because, firstly, they are not independent as we can see from figure~\ref{Fig1}: we can suppose, for example, $\theta_{2}-\theta_{3}=\theta_{1}-\theta_{3}-(\theta_{1}-\theta_{2})$; secondly, we can see that there are off-diagonal terms, as $\widetilde{\partial}_{\mu}\left(\theta_{1}-\theta_{2}\right)\widetilde{\partial}^{\mu}\left(\theta_{1}-\theta_{3}\right)$, in Lagrangian (\ref{2.7}). Thus, in order to find normal oscillations, we must diagonalize Lagrangian (\ref{2.7}). However, due to mathematical cumbersomeness, to find normal oscillations we will proceed from the original Lagrangian (\ref{2.4}).

Before considering the problem of finding the normal frequencies, let us consider ``potential energy'' in the Lagrangian (\ref{2.4}). Substituting the modulus-phase representation (\ref{2.1}) in the Lagrangian~(\ref{2.4}) and assuming $A_{\mu}=0$, we obtain:
\begin{eqnarray}\label{2.8}
  \mathcal{U}&=&a_{1}\left|\Psi_{1}\right|^{2}+a_{2}\left|\Psi_{2}\right|^{2}+a_{3}\left|\Psi_{3}\right|^{2}
    +\frac{b_{1}}{2}\left|\Psi_{1}\right|^{4}+\frac{b_{2}}{2}\left|\Psi_{2}\right|^{4}+\frac{b_{3}}{2}\left|\Psi_{3}\right|^{4}\nonumber\\
  &+&2\epsilon_{12}|\Psi_{1}||\Psi_{2}|\cos(\theta_{1}-\theta_{2})
    +2\epsilon_{13}|\Psi_{1}||\Psi_{3}|\cos(\theta_{1}-\theta_{3})+2\epsilon_{23}|\Psi_{2}||\Psi_{3}|\cos(\theta_{2}-\theta_{3}).\quad
\end{eqnarray}
At $T<T_{c}$, we can consider small variations of the modulus of OP from its equilibrium value: $|\Psi_{1,2,3}|=\Psi_{01,02,03}+\phi_{1,2,3}$, where $|\phi_{1,2,3}|\ll\Psi_{01,02,03}$. Then, $|\Psi|^{2}\approx\Psi_{0}^{2}+2\Psi_{0}\phi+\phi^{2}$,  $|\Psi|^{4}\approx\Psi_{0}^{4}+4\Psi_{0}^{3}\phi+6\Psi_{0}^{2}\phi^{2}$,  $|\Psi_{1}||\Psi_{2}|\approx\Psi_{01}\Psi_{02}+\Psi_{01}\phi_{2}+\Psi_{02}\phi_{1}+\phi_{1}\phi_{2}$. Moreover, we can consider small variations of the phase differences of OP from their equilibrium value:  
\begin{eqnarray}
	\cos\theta_{ik}&=&\cos\big(\theta_{ik}-\theta_{ik}^{0}+\theta_{ik}^{0}\big)
	=\cos\big(\theta_{ik}-\theta_{ik}^{0}\big)\cos\theta_{ik}^{0}-\sin\big(\theta_{ik}-\theta_{ik}^{0}\big)\sin\theta_{ik}^{0} \nonumber\\
	&\approx& \left[1-{\big(\theta_{ik}-\theta_{ik}^{0}\big)^{2}}/{2}\right]\cos\theta_{ik}^{0}-\big(\theta_{ik}-\theta_{ik}^{0}\big)\sin\theta_{ik}^{0},\nonumber
\end{eqnarray} 
where we have introduced the notations $\theta_{i}-\theta_{k}\equiv\theta_{ik}$.
Then, the energy (\ref{2.8}) takes the form:
\begin{eqnarray}\label{2.9}
\mathcal{U}&\approx&\mathcal{U}_{\phi}+\mathcal{U}_{\theta}+\mathcal{U}_{\phi\theta}
+a_{1}\Psi_{01}^{2}+\frac{b_{1}}{2}\Psi_{01}^{4}+a_{2}\Psi_{02}^{2}+\frac{b_{2}}{2}\Psi_{02}^{4}
+a_{3}\Psi_{03}^{2}+\frac{b_{3}}{2}\Psi_{03}^{4}\nonumber\\
&+&2\epsilon_{12}\cos\theta_{12}^{0}\Psi_{01}\Psi_{02}+2\epsilon_{13}\cos\theta_{13}^{0}\Psi_{01}\Psi_{03}
+2\epsilon_{23}\cos\theta_{23}^{0}\Psi_{02}\Psi_{03},
\end{eqnarray}
where the last nine terms determine global potential (as the ``mexican hat''), $\mathcal{U}_{\phi}$ determines a potential for the module excitations $\phi_{1,2,3}$:
\begin{eqnarray}\label{2.10}
\mathcal{U}_{\phi}&=&\phi_{1}^{2}\left(a_{1}+3b_{1}\Psi_{01}^{2}\right)+\phi_{2}^{2}\left(a_{2}+3b_{2}\Psi_{02}^{2}\right)
+\phi_{3}^{2}\left(a_{2}+3b_{3}\Psi_{03}^{2}\right)\nonumber\\
&+&\phi_{1}\phi_{2}2\epsilon_{12}\cos\theta_{12}^{0}  +\phi_{1}\phi_{3}2\epsilon_{13}\cos\theta_{13}^{0}+\phi_{2}\phi_{3}2\epsilon_{23}\cos\theta_{23}^{0}\nonumber\\
&+&2\phi_{1}\left(\epsilon_{12}\cos\theta_{12}^{0}\Psi_{02}+\epsilon_{13}\cos\theta_{13}^{0}\Psi_{03}+a_{1}\Psi_{01}+b_{1}\Psi_{01}^{3}\right)\nonumber\\
&+&2\phi_{2}\left(\epsilon_{12}\cos\theta_{12}^{0}\Psi_{01}+\epsilon_{23}\cos\theta_{23}^{0}\Psi_{03}+a_{2}\Psi_{02}+b_{2}\Psi_{02}^{3}\right)\nonumber\\
&+&2\phi_{3}\left(\epsilon_{13}\cos\theta_{13}^{0}\Psi_{01}+\epsilon_{23}\cos\theta_{23}^{0}\Psi_{02}+a_{3}\Psi_{03}+b_{3}\Psi_{03}^{3}\right).
\end{eqnarray}
The terms at $\phi_{1,2,3}$ should be zero, then
\begin{eqnarray}\label{2.11}
  \epsilon_{12}\cos\theta_{12}^{0}\Psi_{02}+\epsilon_{13}\cos\theta_{13}^{0}\Psi_{03}+a_{1}\Psi_{01}+b_{1}\Psi_{01}^{3}=0, \nonumber\\
  \epsilon_{12}\cos\theta_{12}^{0}\Psi_{01}+\epsilon_{23}\cos\theta_{23}^{0}\Psi_{03}+a_{2}\Psi_{02}+b_{2}\Psi_{02}^{3}=0, \nonumber\\
  \epsilon_{13}\cos\theta_{13}^{0}\Psi_{01}+\epsilon_{23}\cos\theta_{23}^{0}\Psi_{02}+a_{3}\Psi_{03}+b_{3}\Psi_{03}^{3}=0, 
\end{eqnarray}
which corresponds to the first three equations in equation~(\ref{1.2b}). $\mathcal{U}_{\theta}$ determines a potential for the phase excitations $\theta_{1,2,3}$:
\begin{eqnarray}\label{2.12}
\mathcal{U}_{\theta}&=&-2\epsilon_{12}\Psi_{01}\Psi_{02}\frac{\big(\theta_{12}-\theta_{12}^{0}\big)^{2}}{2}
    -2\epsilon_{13}\Psi_{01}\Psi_{03}\frac{\big(\theta_{13}-\theta_{13}^{0}\big)^{2}}{2}
    -2\epsilon_{23}\Psi_{02}\Psi_{03}\frac{\big(\theta_{23}-\theta_{23}^{0}\big)^{2}}{2}\nonumber\\
    &-&2\epsilon_{12}\Psi_{01}\Psi_{02}\big(\theta_{12}-\theta_{12}^{0}\big)\sin\theta_{12}^{0}
    -2\epsilon_{13}\Psi_{01}\Psi_{03}\big(\theta_{13}-\theta_{13}^{0}\big)\sin\theta_{13}^{0}\nonumber\\
    &-&2\epsilon_{23}\Psi_{02}\Psi_{03}\big(\theta_{23}-\theta_{23}^{0}\big)\sin\theta_{23}^{0}.
\end{eqnarray}
For the linear terms $(\theta_{ij}-\theta_{ij}^{0})$ not to affect the equations of motion, the following condition must be satisfied:
\begin{eqnarray}\label{2.13}
  \epsilon_{12}\Psi_{02}\sin\theta_{12}^{0}+\epsilon_{13}\Psi_{03}\sin\theta_{13}^{0}=0,\nonumber\\
  \epsilon_{12}\Psi_{01}\sin\theta_{12}^{0}+\epsilon_{23}\Psi_{03}\sin\theta_{32}^{0}=0, \nonumber\\
  \epsilon_{13}\Psi_{01}\sin\theta_{13}^{0}+\epsilon_{23}\Psi_{02}\sin\theta_{23}^{0}=0,
\end{eqnarray}
which corresponds to the second three equations in equation~(\ref{1.2b}). $\mathcal{U}_{\phi\theta}$ determines the interaction between the module excitations and the phase excitations:
\begin{eqnarray}\label{2.14}
\mathcal{U}_{\phi\theta}&=&-\phi_{1}\phi_{2}\epsilon_{12}
\left[\big(\theta_{12}-\theta_{12}^{0}\big)^{2}\cos\theta_{12}^{0}+2\big(\theta_{12}-\theta_{12}^{0}\big)\sin\theta_{12}^{0}\right]\nonumber\\ 
&-&\phi_{1}\phi_{3}\epsilon_{13}
\left[\big(\theta_{13}-\theta_{13}^{0}\big)^{2}\cos\theta_{13}^{0}+2\big(\theta_{13}-\theta_{13}^{0}\big)\sin\theta_{13}^{0}\right]\nonumber\\
&-&\phi_{2}\phi_{3}\epsilon_{23}
\left[\big(\theta_{23}-\theta_{23}^{0}\big)^{2}\cos\theta_{23}^{0}+2\big(\theta_{23}-\theta_{23}^{0}\big)\sin\theta_{23}^{0}\right]\nonumber\\
&-&\phi_{1}\left[\big(\theta_{12}-\theta_{12}^{0}\big)^{2}\epsilon_{12}\cos\theta_{12}^{0}\Psi_{02}
+\big(\theta_{13}-\theta_{13}^{0}\big)^{2}\epsilon_{13}\cos\theta_{13}^{0}\Psi_{03}\right]\nonumber\\
&-&\phi_{2}\left[\big(\theta_{12}-\theta_{12}^{0}\big)^{2}\epsilon_{12}\cos\theta_{12}^{0}\Psi_{01}
+\big(\theta_{23}-\theta_{23}^{0}\big)^{2}\epsilon_{23}\cos\theta_{23}^{0}\Psi_{03}\right]\nonumber\\
&-&\phi_{3}\left[\big(\theta_{13}-\theta_{13}^{0}\big)^{2}\epsilon_{13}\cos\theta_{13}^{0}\Psi_{01}
+\big(\theta_{23}-\theta_{23}^{0}\big)\epsilon_{23}\sin\theta_{23}^{0}\Psi_{02}\right]\nonumber\\
&-&2\phi_{1}\left[\big(\theta_{12}-\theta_{12}^{0}\big)\epsilon_{12}\sin\theta_{12}^{0}\Psi_{02}
+\big(\theta_{13}-\theta_{13}^{0}\big)\epsilon_{13}\sin\theta_{13}^{0}\Psi_{03}\right]\nonumber\\
&-&2\phi_{2}\left[\big(\theta_{12}-\theta_{12}^{0}\big)\epsilon_{12}\sin\theta_{12}^{0}\Psi_{01}
+\big(\theta_{23}-\theta_{23}^{0}\big)\epsilon_{23}\sin\theta_{23}^{0}\Psi_{03}\right]\nonumber\\
&-&2\phi_{3}\left[\big(\theta_{13}-\theta_{13}^{0}\big)\epsilon_{13}\sin\theta_{13}^{0}\Psi_{01}
+\big(\theta_{23}-\theta_{23}^{0}\big)\epsilon_{23}\sin\theta_{23}^{0}\Psi_{02}\right].
\end{eqnarray}
We can see that the first six terms are of the third $\phi_{i}\phi_{k}(\theta_{ik}-\theta_{ik}^{0})$,  $\phi_{i}(\theta_{ij}-\theta_{ij}^{0})^{2}$ and the forth $\phi_{i}\phi_{k}(\theta_{ik}-\theta_{ik}^{0})^{2}$ order. Hence, they can be neglected. At the same time, the last three terms are of the second order $\phi_{i}(\theta_{ik}-\theta_{ik}^{0})$.  In the case $\epsilon_{12}\epsilon_{13}\epsilon_{23}<0$, we have all $\theta_{ik}^{0}=0$ or $\piup$, that is $\sin\theta_{ik}^{0}=0$, hence the oscillations of the amplitudes and of the phases are not hybridized in this case. Thus, the Goldstone and the Higgs modes are hybridized in the case $\epsilon_{12}\epsilon_{13}\epsilon_{23}>0$ only, that is the phase-amplitude mode can take place~\cite{stanev2,babaev1}.

The drag terms cause the analogous situation:
\begin{eqnarray}\label{2.15}
&&\left(\widetilde{\partial}_{\mu}\Psi_{i}\widetilde{\partial}^{\mu}\Psi_{k}^{+}
+\widetilde{\partial}^{\mu}\Psi_{i}^{+}\widetilde{\partial}_{\mu}\Psi_{k}\right)\nonumber\\
&&\approx\left[\left(\widetilde{\partial}_{\mu}\phi_{i}\widetilde{\partial}^{\mu}\phi_{k}
    +\widetilde{\partial}_{\mu}\phi_{k}\widetilde{\partial}^{\mu}\phi_{i}\right)+
  \Psi_{01}\Psi_{02}(\widetilde{\partial}_{\mu}\theta_{i}\widetilde{\partial}^{\mu}\theta_{k}+
  \widetilde{\partial}_{\mu}\theta_{k}\widetilde{\partial}^{\mu}\theta_{i})\right]\cos\theta_{ik}^{0} \nonumber\\
   &&-\left[\Psi_{0i}\left(\widetilde{\partial}_{\mu}\phi_{i}\widetilde{\partial}^{\mu}\theta_{k}
    +\widetilde{\partial}_{\mu}\theta_{k}\widetilde{\partial}^{\mu}\phi_{i}\right)-
  \Psi_{0k}(\widetilde{\partial}_{\mu}\theta_{i}\widetilde{\partial}^{\mu}\phi_{k}+
  \widetilde{\partial}_{\mu}\phi_{k}\widetilde{\partial}^{\mu}\theta_{i})\right]\sin\theta_{ik}^{0}.
\end{eqnarray}
We can see that in the case $\epsilon_{12}\epsilon_{13}\epsilon_{23}<0$, the oscillations of the phase $\theta$ and amplitude $\phi$ are not hybridized the same as for potential energy (\ref{2.14}). It should be noted that the phase-amplitude hybridization is absent in two-band superconductors due to this property. In addition, as demonstrated in reference~\cite{pal2}, the effect of the mixing of the oscillations of phases and amplitudes of OP from different bands is essential for a reduced charge carrier density $\mu<\omega_{\rm D}$ ($\mu$ is chemical potential, $\omega_{\rm D}$ is Debye frequency). For a large charge carrier density, the oscillations of phases and amplitudes can be supposed independent.

Accounting of the hybridization results in the dispersion equation of the sixth order, instead of two equations (\ref{3.6a}) and (\ref{4.10}) of the third order. The sixth order equation cannot be solved analytically. In order to obtain an analytical spectrum of quasiparticles, we are forced to use the decoupling of correlations. As will be demonstrated below, the spectrum of collective excitations is determined not only by the coefficients of the proximity effect $\epsilon_{ik}$, but also by the coefficients of the drag effect $\eta_{ik}$. As in two-band superconductors, the properties of the Higgs modes at $T=T_{c}$ force us to regard the coefficients~$\eta_{ik}$  in such way that it leaves only the common mode Higgs and Goldstone oscillations. Special choice of the coefficients~$\eta_{ik}$, which eliminates the spectrum branches with anti-phase oscillations, is \emph{the same} both for the Leggett modes and for the Higgs modes if we neglect their hybridization, and regardless of the sign of~$\epsilon_{12}\epsilon_{13}\epsilon_{23}$. Hence, at the first stage, we can consider the normal oscillations without the phase-amplitude hybridization. Accounting of the phase-amplitude hybridization \emph{requires special consideration}.



\subsection{Goldstone modes}

\begin{figure}[h]
\centerline{\includegraphics[width=0.5\textwidth]{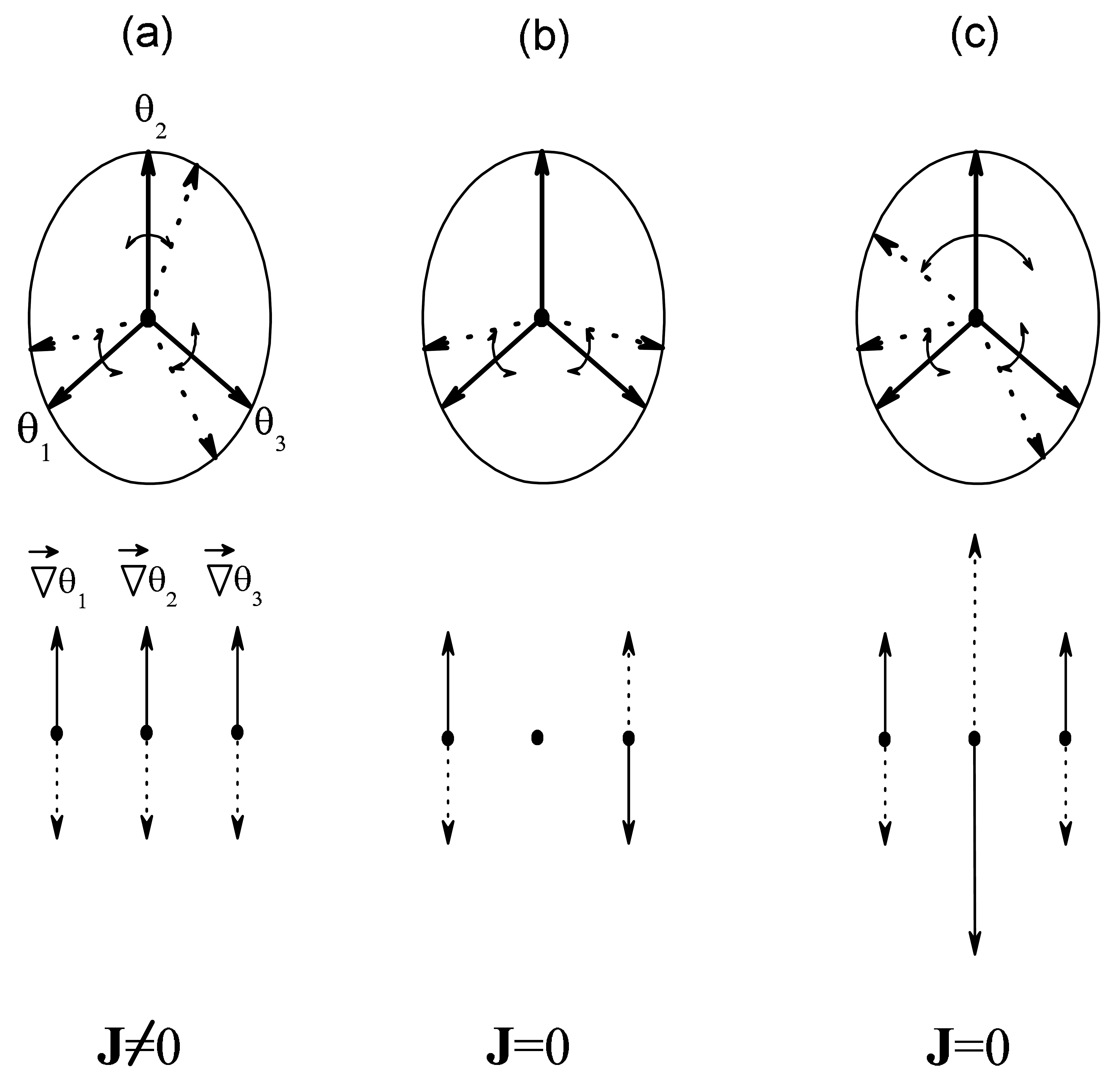}}
\caption{Normal oscillations of the phases $\theta_{1}$, $\theta_{2}$, $\theta_{3}$ in a symmetrical three-band system $|\Psi_{1}|=|\Psi_{2}|=|\Psi_{3}|$, $m_{1}=m_{2}=m_{3}$ with repulsive interband interactions $\epsilon_{12}=\epsilon_{13}=\epsilon_{23}>0$ in the case of the absence of the drag effect $\eta_{12}=\eta_{13}=\eta_{23}=0$. (a) Common phase oscillations with acoustic spectrum (\ref{3.8}), which are accompanied by nonzero current $\mathbf{J}=\frac{\hbar e}{m_{1}}|\Psi_{1}|^{2}\nabla\theta_{1}+\frac{\hbar e}{m_{2}}|\Psi_{2}|^{2}\nabla\theta_{2}+\frac{\hbar e}{m_{3}}|\Psi_{3}|^{2}\nabla\theta_{3}\neq 0$. (b, c) Anti-phase oscillations with the massive spectrum (\ref{3.10}), which are not accompanied by the current, i.e. $\mathbf{J}=0$.}
\label{Fig3}
\end{figure}

Let us consider the movement of the phases only. Using the modulus-phase representation (\ref{2.1}) and assuming $|\Psi_{1,2,3}|=\mathrm{const}$ and $A_{\mu}=0$, the Lagrangian (\ref{2.4}) takes the form:
\begin{eqnarray}\label{3.1}
    \mathcal{L}&=&
    \frac{\hbar^{2}}{4m_{1}}|\Psi_{1}|^{2}\widetilde{\partial}_{\mu}\theta_{1}\widetilde{\partial}^{\mu}\theta_{1}
    +\frac{\hbar^{2}}{4m_{2}}|\Psi_{2}|^{2}\widetilde{\partial}_{\mu}\theta_{2}\widetilde{\partial}^{\mu}\theta_{2}
    +\frac{\hbar^{3}}{4m_{3}}|\Psi_{3}|^{2}\widetilde{\partial}_{\mu}\theta_{3}\widetilde{\partial}^{\mu}\theta_{3}\nonumber\\
    &+&\frac{\hbar^{2}}{4}\eta_{12}|\Psi_{1}||\Psi_{2}|
    \big(\widetilde{\partial}_{\mu}\theta_{1}\widetilde{\partial}^{\mu}\theta_{2}+
    \widetilde{\partial}_{\mu}\theta_{2}\widetilde{\partial}^{\mu}\theta_{1}\big)\cos(\theta_{1}-\theta_{2})\nonumber\\
    &+&\frac{\hbar^{2}}{4}\eta_{13}|\Psi_{1}||\Psi_{3}|
    \big(\widetilde{\partial}_{\mu}\theta_{1}\widetilde{\partial}^{\mu}\theta_{3}+
    \widetilde{\partial}_{\mu}\theta_{3}\widetilde{\partial}^{\mu}\theta_{1}\big)\cos(\theta_{1}-\theta_{3})\nonumber\\
    &+&\frac{\hbar^{2}}{4}\eta_{23}|\Psi_{2}||\Psi_{3}|
    \big(\widetilde{\partial}_{\mu}\theta_{2}\widetilde{\partial}^{\mu}\theta_{3}+
    \widetilde{\partial}_{\mu}\theta_{3}\widetilde{\partial}^{\mu}\theta_{2}\big)\cos(\theta_{2}-\theta_{3})\nonumber\\
    &-&2\epsilon_{12}|\Psi_{1}||\Psi_{2}|\cos\left(\theta_{1}-\theta_{2}\right)
    -2\epsilon_{13}|\Psi_{1}||\Psi_{3}|\cos\left(\theta_{1}-\theta_{3}\right)\nonumber\\
    &-&2\epsilon_{23}|\Psi_{2}||\Psi_{3}|\cos\left(\theta_{2}-\theta_{3}\right)
    +\mathcal{L}\left(|\Psi_{1}|,|\Psi_{2}|,|\Psi_{3}|\right).
\end{eqnarray}
Corresponding  Lagrange equation, for example, is
\begin{eqnarray}\label{3.2}
  \widetilde{\partial}_{\mu}\frac{\partial\mathcal{L}}{\partial(\widetilde{\partial}_{\mu}\theta_{1})}
  -\frac{\partial\mathcal{L}}{\partial\theta_{1}}=0\Rightarrow&&
  \frac{\hbar^{2}}{4m_{1}}|\Psi_{1}|^{2}\widetilde{\partial}_{\mu}\widetilde{\partial}^{\mu}\theta_{1}
  +\frac{\hbar^{2}}{4}\eta_{12}|\Psi_{1}||\Psi_{2}|\cos(\theta_{1}-\theta_{2})\widetilde{\partial}_{\mu}\widetilde{\partial}^{\mu}\theta_{2}\nonumber\\
&&+\frac{\hbar^{2}}{4}\eta_{13}|\Psi_{1}||\Psi_{3}|\cos(\theta_{1}-\theta_{3})\widetilde{\partial}_{\mu}\widetilde{\partial}^{\mu}\theta_{3}\nonumber\\
&&-|\Psi_{1}||\Psi_{2}|\epsilon_{12}\sin(\theta_{1}-\theta_{2})-|\Psi_{1}||\Psi_{3}|\epsilon_{13}\sin(\theta_{1}-\theta_{3})=0,\qquad
\end{eqnarray}
where we have omitted nonlinear terms $\widetilde{\partial}_{\mu}\theta\widetilde{\partial}^{\mu}\theta$. The phases can be written in the form of harmonic oscillations:
\begin{eqnarray}\label{3.3}
      \theta_{1}&=&\theta_{1}^{0}+A\re^{\ri(\mathbf{qr}-\omega t)}\equiv\theta_{1}^{0}+A\re^{-\ri q_{\mu}x^{\mu}}, \nonumber\\
      \theta_{2}&=&\theta_{2}^{0}+B\re^{\ri(\mathbf{qr}-\omega t)}\equiv\theta_{2}^{0}+B\re^{-\ri q_{\mu}x^{\mu}}, \nonumber\\
      \theta_{3}&=&\theta_{3}^{0}+C\re^{\ri(\mathbf{qr}-\omega t)}\equiv\theta_{3}^{0}+C\re^{-\ri q_{\mu}x^{\mu}},
\end{eqnarray}
where $q_{\mu}=\left({\omega}/{\upsilon},-\mathbf{q}\right)$,  $x^{\mu}=\left(\upsilon t,\mathbf{r}\right)$, $\theta_{1,2,3}^{0}$ are equilibrium phases. We should linearize equation~(\ref{3.2}) assuming $\cos\theta_{ik}\approx\cos\theta_{ik}^{0}$,  $\sin\theta_{ik}=\sin\big(\theta_{ik}-\theta_{ik}^{0}+\theta_{ik}^{0}\big)
\approx\big(\theta_{ik}-\theta_{ik}^{0}\big)\cos\theta_{ik}^{0}+\sin\theta_{ik}^{0}$, and using the second three equations from equation~(\ref{1.2b}). Then, the linearized equations are
\begin{eqnarray}\label{3.4}
  &&\frac{\hbar^{2}}{4m_{1}}|\Psi_{1}|^{2}\widetilde{\partial}_{\mu}\widetilde{\partial}^{\mu}\theta_{1}+
  \frac{\hbar^{2}}{4}\big[\eta_{12}\cos\theta_{12}^{0}\big]|\Psi_{1}||\Psi_{2}|\widetilde{\partial}_{\mu}\widetilde{\partial}^{\mu}\theta_{2}
  +\frac{\hbar^{2}}{4}\big[\eta_{13}\cos\theta_{13}^{0}\big]|\Psi_{1}||\Psi_{3}|\widetilde{\partial}_{\mu}\widetilde{\partial}^{\mu}\theta_{3}\nonumber\\
&&{}-|\Psi_{1}||\Psi_{2}|\big[\epsilon_{12}\cos\theta_{12}^{0}\big]\big(\theta_{12}-\theta_{12}^{0}\big)
-|\Psi_{1}||\Psi_{3}|\big[\epsilon_{13}\cos\theta_{13}^{0}\big]\big(\theta_{13}-\theta_{13}^{0}\big)=0,\nonumber\\
&&\frac{\hbar^{2}}{4m_{2}}|\Psi_{2}|^{2}\widetilde{\partial}_{\mu}\widetilde{\partial}^{\mu}\theta_{2}+
  \frac{\hbar^{2}}{4}\big[\eta_{12}\cos\theta_{12}^{0}\big]|\Psi_{1}||\Psi_{2}|\widetilde{\partial}_{\mu}\widetilde{\partial}^{\mu}\theta_{1}
  +\frac{\hbar^{2}}{4}\big[\eta_{23}\cos\theta_{23}^{0}\big]|\Psi_{2}||\Psi_{3}|\widetilde{\partial}_{\mu}\widetilde{\partial}^{\mu}\theta_{3}\nonumber\\
&&{}+|\Psi_{1}||\Psi_{2}|\big[\epsilon_{12}\cos\theta_{12}^{0}\big]\big(\theta_{12}-\theta_{12}^{0}\big)
-|\Psi_{2}||\Psi_{3}|\big[\epsilon_{23}\cos\theta_{23}^{0}\big]\big(\theta_{23}-\theta_{23}^{0}\big)=0,\nonumber\\
&&\frac{\hbar^{2}}{4m_{3}}|\Psi_{3}|^{2}\widetilde{\partial}_{\mu}\widetilde{\partial}^{\mu}\theta_{3}+
  \frac{\hbar^{2}}{4}\big[\eta_{13}\cos\theta_{13}^{0}\big]|\Psi_{1}||\Psi_{3}|\widetilde{\partial}_{\mu}\widetilde{\partial}^{\mu}\theta_{1}
  +\frac{\hbar^{2}}{4}\big[\eta_{23}\cos\theta_{23}^{0}\big]|\Psi_{2}||\Psi_{3}|\widetilde{\partial}_{\mu}\widetilde{\partial}^{\mu}\theta_{2}\nonumber\\
&&{}+|\Psi_{1}||\Psi_{3}|\big[\epsilon_{13}\cos\theta_{13}^{0}\big]\big(\theta_{13}-\theta_{13}^{0}\big)
+|\Psi_{2}||\Psi_{3}|\big[\epsilon_{23}\cos\theta_{23}^{0}\big]\big(\theta_{23}-\theta_{23}^{0}\big)=0.
\end{eqnarray}
Substituting equation~(\ref{3.3}) in equation~(\ref{3.4}), we obtain equations for the amplitudes $A,B,C$:
\begin{eqnarray}\label{3.5}
&&A\left(-\frac{|\Psi_{2}|}{|\Psi_{1}|}\epsilon_{12}\cos\theta_{12}^{0}-
\frac{|\Psi_{3}|}{|\Psi_{1}|}\epsilon_{13}\cos\theta_{13}^{0}-q_{\mu}q^{\mu}\frac{\hbar^{2}}{4m_{1}}\right)
  +B\frac{|\Psi_{2}|}{|\Psi_{1}|}\left(\epsilon_{12}\cos\theta_{12}^{0}-q_{\mu}q^{\mu}\frac{\hbar^{2}}{4}\eta_{12}\cos\theta_{12}^{0}\right)\nonumber\\
  &&{}+C\frac{|\Psi_{3}|}{|\Psi_{1}|}\left(\epsilon_{13}\cos\theta_{13}^{0}
  -q_{\mu}q^{\mu}\frac{\hbar^{2}}{4}\eta_{13}\cos\theta_{13}^{0}\right)=0,\nonumber\\
&&A\frac{|\Psi_{1}|}{|\Psi_{2}|}\left(\epsilon_{12}\cos\theta_{12}^{0}-q_{\mu}q^{\mu}\frac{\hbar^{2}}{4}\eta_{12}\cos\theta_{12}^{0}\right)
  +B\left(-\frac{|\Psi_{1}|}{|\Psi_{2}|}\epsilon_{12}\cos\theta_{12}^{0}-
      \frac{|\Psi_{3}|}{|\Psi_{2}|}\epsilon_{23}\cos\theta_{23}^{0}-q_{\mu}q^{\mu}\frac{\hbar^{2}}{4m_{2}}\right)\nonumber\\
  &&{}+C\frac{|\Psi_{3}|}{|\Psi_{2}|}\left(\epsilon_{23}\cos\theta_{23}^{0}
  -q_{\mu}q^{\mu}\frac{\hbar^{2}}{4}\eta_{23}\cos\theta_{23}^{0}\right)=0,\nonumber\\
&&A\frac{|\Psi_{1}|}{|\Psi_{3}|}\left(\epsilon_{13}\cos\theta_{13}^{0}-q_{\mu}q^{\mu}\frac{\hbar^{2}}{4}\eta_{13}\cos\theta_{13}^{0}\right)
  +B\frac{|\Psi_{2}|}{|\Psi_{3}|}\left(\epsilon_{23}\cos\theta_{23}^{0}
  -q_{\mu}q^{\mu}\frac{\hbar^{2}}{4}\eta_{23}\cos\theta_{23}^{0}\right)\nonumber\\
  &&{}+C\left(-\frac{|\Psi_{1}|}{|\Psi_{3}|}\epsilon_{13}\cos\theta_{13}^{0}-
      \frac{|\Psi_{2}|}{|\Psi_{3}|}\epsilon_{23}\cos\theta_{23}^{0}-q_{\mu}q^{\mu}\frac{\hbar^{2}}{4m_{3}}\right)=0.
\end{eqnarray}
Equating the determinant of the system to zero (\ref{3.5}), we find a dispersion equation:
\begin{equation}\label{3.6a}
\left(q_{\mu}q^{\mu}\right)^{3}a+\left(q_{\mu}q^{\mu}\right)^{2}b+\left(q_{\mu}q^{\mu}\right)c=0,
\end{equation}
where
\begin{eqnarray}\label{3.6}
a&=&\left(\frac{\hbar^{2}}{4}\right)^{3}\left[\frac{1}{m_{1}m_{2}m_{3}}
+2\widetilde{\eta}_{12}\widetilde{\eta}_{13}\widetilde{\eta}_{23}
-\frac{\widetilde{\eta}_{12}^{2}}{m_{3}}-\frac{\widetilde{\eta}_{13}^{2}}{m_{2}}-\frac{\widetilde{\eta}_{23}^{2}}{m_{1}}\right],\nonumber\\
b&=&\left(\frac{\hbar^{2}}{4}\right)^{2}\bigg[\left(\frac{|\Psi_{1}|}{|\Psi_{3}|}\widetilde{\epsilon}_{13}+
\frac{|\Psi_{2}|}{|\Psi_{3}|}\widetilde{\epsilon}_{23}\right)\left(\frac{1}{m_{1}m_{2}}-\eta_{12}^{2}\right)
+\left(\frac{|\Psi_{1}|}{|\Psi_{2}|}\widetilde{\epsilon}_{12}+
\frac{|\Psi_{3}|}{|\Psi_{2}|}\widetilde{\epsilon}_{23}\right)\left(\frac{1}{m_{1}m_{3}}-\eta_{13}^{2}\right)\nonumber\\
&&{}+\left(\frac{|\Psi_{2}|}{|\Psi_{1}|}\widetilde{\epsilon}_{12}+
\frac{|\Psi_{3}|}{|\Psi_{1}|}\widetilde{\epsilon}_{13}\right)\left(\frac{1}{m_{2}m_{3}}-\eta_{23}^{2}\right)
-2\widetilde{\epsilon}_{12}\left(\frac{\widetilde{\eta}_{12}}{m_{3}}-\widetilde{\eta}_{13}\widetilde{\eta}_{23}\right)
-2\widetilde{\epsilon}_{13}\left(\frac{\widetilde{\eta}_{13}}{m_{2}}-\widetilde{\eta}_{12}\widetilde{\eta}_{23}\right)\nonumber\\
&&{}-2\widetilde{\epsilon}_{23}\left(\frac{\widetilde{\eta}_{23}}{m_{1}}-\widetilde{\eta}_{12}\widetilde{\eta}_{13}\right)\bigg],\nonumber\\
c&=&\frac{\hbar^{2}}{4}
\bigg[\left(\frac{|\Psi_{1}|}{|\Psi_{2}|}\widetilde{\epsilon}_{12}+\frac{|\Psi_{3}|}{|\Psi_{2}|}\widetilde{\epsilon}_{23}\right)
\left(\frac{|\Psi_{1}|}{|\Psi_{3}|}\widetilde{\epsilon}_{13}+\frac{|\Psi_{2}|}{|\Psi_{3}|}\widetilde{\epsilon}_{23}\right)\frac{1}{m_{1}}
+\left(\frac{|\Psi_{2}|}{|\Psi_{1}|}\widetilde{\epsilon}_{12}+\frac{|\Psi_{3}|}{|\Psi_{1}|}\widetilde{\epsilon}_{13}\right)\nonumber\\
&&{}\times
\left(\frac{|\Psi_{1}|}{|\Psi_{3}|}\widetilde{\epsilon}_{13}+\frac{|\Psi_{2}|}{|\Psi_{3}|}\widetilde{\epsilon}_{23}\right)\frac{1}{m_{2}}
+\left(\frac{|\Psi_{2}|}{|\Psi_{1}|}\widetilde{\epsilon}_{12}+\frac{|\Psi_{3}|}{|\Psi_{1}|}\widetilde{\epsilon}_{13}\right)
\left(\frac{|\Psi_{1}|}{|\Psi_{2}|}\widetilde{\epsilon}_{12}+\frac{|\Psi_{3}|}{|\Psi_{2}|}\widetilde{\epsilon}_{23}\right)\frac{1}{m_{3}}
+2\widetilde{\epsilon}_{13}\widetilde{\epsilon}_{23}\widetilde{\eta}_{12} \nonumber\\
&&{}+2\widetilde{\epsilon}_{12}\widetilde{\epsilon}_{23}\widetilde{\eta}_{13}
+2\widetilde{\epsilon}_{12}\widetilde{\epsilon}_{13}\widetilde{\eta}_{23}
-\widetilde{\epsilon}_{12}^{2}\frac{1}{m_{3}}+2\widetilde{\epsilon}_{12}\widetilde{\eta}_{12}
\left(\frac{|\Psi_{1}|}{|\Psi_{3}|}\widetilde{\epsilon}_{13}+\frac{|\Psi_{2}|}{|\Psi_{3}|}\widetilde{\epsilon}_{23}\right)
-\widetilde{\epsilon}_{13}^{2}\frac{1}{m_{2}}\nonumber\\
&&{}+2\widetilde{\epsilon}_{13}\widetilde{\eta}_{13}
\left(\frac{|\Psi_{1}|}{|\Psi_{2}|}\widetilde{\epsilon}_{12}+\frac{|\Psi_{3}|}{|\Psi_{2}|}\widetilde{\epsilon}_{23}\right)
-\widetilde{\epsilon}_{23}^{2}\frac{1}{m_{1}}+2\widetilde{\epsilon}_{23}\widetilde{\eta}_{23}
\left(\frac{|\Psi_{2}|}{|\Psi_{1}|}\widetilde{\epsilon}_{21}+\frac{|\Psi_{3}|}{|\Psi_{1}|}\widetilde{\epsilon}_{13}\right)\bigg],
\end{eqnarray}
and we denoted:
\begin{equation}\label{3.7}
  \widetilde{\epsilon}_{ik}\equiv\epsilon_{ik}\cos\theta_{ik}^{0},\qquad\widetilde{\eta}_{ik}\equiv\eta_{ik}\cos\theta_{ik}^{0}.
\end{equation}
From equation~(\ref{3.6a}) we can see that one of dispersion relations is
\begin{equation}\label{3.8}
    q_{\mu}q^{\mu}=0\Rightarrow\omega^{2}=q^{2}\upsilon^{2},
\end{equation}
wherein $A=B=C$. Thus, this mode represents the common mode oscillations, as Goldstone mode in single-band superconductors. \emph{There are other oscillation modes with such spectra, that $q_{\mu}q^{\mu}=(-b\pm\sqrt{b^{2}-4ac})/2a\neq 0$, i.e., two massive modes. These modes are analogous to the Leggett mode in two-band superconductors}~\cite{grig3} and correspond to the results of references~\cite{pal1,pal2,ota} for the phase oscillations in three-band superconductors. It should be noted that if we assume all $\epsilon_{ik}=0$, then $b=c=0$ and the dispersion equation will be $a(q_{\mu}q^{\mu})^{3}=0$. That is, we obtain independent common mode oscillations in each band. Let us consider a symmetrical three-band system $|\Psi_{1}|=|\Psi_{2}|=|\Psi_{3}|$, $m_{1}=m_{2}=m_{3}\equiv m$, $\epsilon_{12}=\epsilon_{13}=\epsilon_{23}\equiv\epsilon$ in the case of the absence of the drag effect $\eta_{12}=\eta_{13}=\eta_{23}=0$. Then, massive modes have the same spectrum:
\begin{equation}\label{3.10}
    q_{\mu}q^{\mu}=-\frac{12}{\hbar^{2}}m\widetilde{\epsilon},
\end{equation}
where $\widetilde{\epsilon}<0$. Amplitudes of these modes relate as $A=-C$, $B=0$ and $A=C$, $B=-(A+C)$, so that current $\mathbf{J}=\frac{\hbar e}{m_{1}}|\Psi_{1}|^{2}\nabla\theta_{1}+\frac{\hbar e}{m_{2}}|\Psi_{2}|^{2}\nabla\theta_{2}+\frac{\hbar e}{m_{3}}|\Psi_{3}|^{2}\nabla\theta_{3}$ is $\mathbf{J}\neq 0$ for the acoustic mode (\ref{3.8}) and $\mathbf{J}=0$ for the massive modes (\ref{3.10}). These three Goldstone modes are shown in figure~\ref{Fig3}.

It is not difficult to see that if we assume
\begin{equation}\label{3.9}
  \widetilde{\eta}_{12}=\frac{1}{\sqrt{m_{1}m_{2}}},\quad\widetilde{\eta}_{13}=\frac{1}{\sqrt{m_{1}m_{3}}},\quad
  \widetilde{\eta}_{23}=\frac{1}{\sqrt{m_{2}m_{3}}},
\end{equation}
then, $a=b=0$. Hence, the common mode oscillations (\ref{3.8}) remain only.

\subsection{Higgs modes}


Let us consider the movement of the modules only (that is, assuming $\theta_{1,2,3}=\theta_{1,2,3}^{0}$), then, the Lagrangian~(\ref{2.4}) takes the form (when $A_{\mu}=0$):
\begin{eqnarray}\label{4.1}
    \mathcal{L}&=&\frac{\hbar^{2}}{4m_{1}}\widetilde{\partial}_{\mu}|\Psi_{1}|\widetilde{\partial}^{\mu}|\Psi_{1}|
    +\frac{\hbar^{2}}{4m_{2}}\widetilde{\partial}_{\mu}|\Psi_{2}|\widetilde{\partial}^{\mu}|\Psi_{2}|
    +\frac{\hbar^{2}}{4m_{3}}\widetilde{\partial}_{\mu}|\Psi_{3}|\widetilde{\partial}^{\mu}|\Psi_{3}|\nonumber\\
    &+&\frac{\hbar^{2}}{4}\widetilde{\eta}_{12}\left(\widetilde{\partial}_{\mu}|\Psi_{1}|\widetilde{\partial}^{\mu}|\Psi_{2}|
    +\widetilde{\partial}_{\mu}|\Psi_{2}|\widetilde{\partial}^{\mu}|\Psi_{1}|\right)
    +\frac{\hbar^{2}}{4}\widetilde{\eta}_{13}\left(\widetilde{\partial}_{\mu}|\Psi_{1}|\widetilde{\partial}^{\mu}|\Psi_{3}|
    +\widetilde{\partial}_{\mu}|\Psi_{3}|\widetilde{\partial}^{\mu}|\Psi_{1}|\right) \nonumber\\
    &+&\frac{\hbar^{2}}{4}\widetilde{\eta}_{23}\left(\widetilde{\partial}_{\mu}|\Psi_{2}|\widetilde{\partial}^{\mu}|\Psi_{3}|
    +\widetilde{\partial}_{\mu}|\Psi_{3}|\widetilde{\partial}^{\mu}|\Psi_{2}|\right)
    -a_{1}\left|\Psi_{1}\right|^{2}-\frac{b_{1}}{2}\left|\Psi_{1}\right|^{4}
    -a_{2}\left|\Psi_{2}\right|^{2}-\frac{b_{2}}{2}\left|\Psi_{2}\right|^{4} \nonumber\\
    &-&a_{3}\left|\Psi_{3}\right|^{2}-\frac{b_{3}}{2}\left|\Psi_{3}\right|^{4}
    -2\widetilde{\epsilon}_{12}|\Psi_{1}||\Psi_{2}|-2\widetilde{\epsilon}_{13}|\Psi_{1}||\Psi_{3}|
    -2\widetilde{\epsilon}_{23}|\Psi_{2}||\Psi_{3}|.
\end{eqnarray}
At $T<T_{c}$, we can consider small variations of the modulus of OP from its equilibrium value: $|\Psi_{1,2,3}|=\Psi_{01,02,03}+\phi_{1,2,3}$, where $|\phi_{1,2,3}|\ll\Psi_{01,02,03}$. Then, $|\Psi|^{2}\approx\Psi_{0}^{2}+2\Psi_{0}\phi+\phi^{2}$,  $|\Psi|^{4}\approx\Psi_{0}^{4}+4\Psi_{0}^{3}\phi+6\Psi_{0}^{2}\phi^{2}$,  $|\Psi_{1}||\Psi_{2}|\approx\Psi_{01}\Psi_{02}+\Psi_{01}\phi_{2}+\Psi_{02}\phi_{1}+\phi_{1}\phi_{2}$, and Lagrangian (\ref{4.1}) takes the form:
\begin{eqnarray}\label{4.2}
    \mathcal{L}&=&\frac{\hbar^{2}}{4m_{1}}\widetilde{\partial}_{\mu}\phi_{1}\widetilde{\partial}^{\mu}\phi_{1}
    +\frac{\hbar^{2}}{4m_{2}}\widetilde{\partial}_{\mu}\phi_{2}\widetilde{\partial}^{\mu}\phi_{2}
    +\frac{\hbar^{2}}{4m_{3}}\widetilde{\partial}_{\mu}\phi_{3}\widetilde{\partial}^{\mu}\phi_{3}
    +\frac{\hbar^{2}}{4}\widetilde{\eta}_{12}\left(\widetilde{\partial}_{\mu}\phi_{1}\widetilde{\partial}^{\mu}\phi_{2}
    +\widetilde{\partial}_{\mu}\phi_{2}\widetilde{\partial}^{\mu}\phi_{1}\right)\nonumber\\
    &+&\frac{\hbar^{2}}{4}\widetilde{\eta}_{13}\left(\widetilde{\partial}_{\mu}\phi_{1}\widetilde{\partial}^{\mu}\phi_{3}
    +\widetilde{\partial}_{\mu}\phi_{3}\widetilde{\partial}^{\mu}\phi_{1}\right)
    +\frac{\hbar^{2}}{4}\widetilde{\eta}_{23}\left(\widetilde{\partial}_{\mu}\phi_{2}\widetilde{\partial}^{\mu}\phi_{3}
    +\widetilde{\partial}_{\mu}\phi_{3}\widetilde{\partial}^{\mu}\phi_{2}\right)
    -\phi_{1}^{2}\left(a_{1}+3b_{1}\Psi_{01}^{2}\right) \nonumber\\
    &-&\phi_{2}^{2}\left(a_{2}+3b_{2}\Psi_{02}^{2}\right)
    -\phi_{3}^{2}\left(a_{2}+3b_{3}\Psi_{03}^{2}\right)-2\widetilde{\epsilon}_{12}\phi_{1}\phi_{2}
    -2\widetilde{\epsilon}_{13}\phi_{1}\phi_{3}-2\widetilde{\epsilon}_{23}\phi_{2}\phi_{3}\nonumber\\
    &-&2\phi_{1}\left(\widetilde{\epsilon}_{12}\Psi_{02}+\widetilde{\epsilon}_{13}\Psi_{03}+a_{1}\Psi_{01}+b_{1}\Psi_{01}^{3}\right)
    -2\phi_{2}\left(\widetilde{\epsilon}_{12}\Psi_{01}+\widetilde{\epsilon}_{23}\Psi_{03}+a_{2}\Psi_{02}+b_{2}\Psi_{02}^{3}\right)\nonumber\\
    &-&2\phi_{3}\left(\widetilde{\epsilon}_{13}\Psi_{01}+\widetilde{\epsilon}_{23}\Psi_{02}+a_{3}\Psi_{03}+b_{3}\Psi_{03}^{3}\right)
    -a_{1}\Psi_{01}^{2}-\frac{b_{1}}{2}\Psi_{01}^{4}-a_{2}\Psi_{02}^{2}-\frac{b_{2}}{2}\Psi_{02}^{4}\nonumber\\
    &-&a_{3}\Psi_{03}^{2}-\frac{b_{3}}{2}\Psi_{03}^{4}
    -2\widetilde{\epsilon}_{12}\Psi_{01}\Psi_{02}-2\widetilde{\epsilon}_{13}\Psi_{01}\Psi_{03}-2\widetilde{\epsilon}_{23}\Psi_{02}\Psi_{03}.
\end{eqnarray}
The last nine terms can be omitted as a constant. The terms at $\phi_{1,2,3}$ should be zero, then
\begin{eqnarray}\label{4.3}
  \widetilde{\epsilon}_{12}\Psi_{02}+\widetilde{\epsilon}_{13}\Psi_{03}+a_{1}\Psi_{01}+b_{1}\Psi_{01}^{3}=0, \nonumber\\
  \widetilde{\epsilon}_{12}\Psi_{01}+\widetilde{\epsilon}_{23}\Psi_{03}+a_{2}\Psi_{02}+b_{2}\Psi_{02}^{3}=0, \nonumber\\
  \widetilde{\epsilon}_{13}\Psi_{01}+\widetilde{\epsilon}_{23}\Psi_{02}+a_{3}\Psi_{03}+b_{3}\Psi_{03}^{3}=0,
\end{eqnarray}
which corresponds to the first three equations in equation~(\ref{1.2b}). At $T>T_{c1}$, $T_{c2}$, $T_{c3}$, we have $a_{1,2,3}>0$ and equation~(\ref{1.3}) in $T=T_{c}$, at $T<T_{c1}$, $T_{c2}$, $T_{c3}$ we have $a_{1,2,3}<0$. At $T\ll T_{c1}$, $T_{c2}$, $T_{c3}$ in the case of the \emph{weak interband coupling} $\epsilon_{ik}^{2}\ll a_{i}a_{k}$, it is not difficult to obtain from equation~(\ref{4.3}):
\begin{eqnarray}\label{4.4}
 \Psi_{01}&=&\sqrt{\frac{|a_{1}|}{b_{1}}}
\left(1-\frac{\widetilde{\epsilon}_{12}}{2\sqrt{|a_{1}||a_{2}|}}\sqrt{\frac{b_{1}}{b_{2}}}\frac{|a_{2}|}{|a_{1}|}
-\frac{\widetilde{\epsilon}_{13}}{2\sqrt{|a_{1}||a_{3}|}}\sqrt{\frac{b_{1}}{b_{3}}}\frac{|a_{3}|}{|a_{1}|}\right)\approx\sqrt{\frac{|a_{1}|}{b_{1}}},\nonumber\\
 \Psi_{02}&=&\sqrt{\frac{|a_{2}|}{b_{2}}}
\left(1-\frac{\widetilde{\epsilon}_{12}}{2\sqrt{|a_{2}||a_{1}|}}\sqrt{\frac{b_{2}}{b_{1}}}\frac{|a_{1}|}{|a_{2}|}
-\frac{\widetilde{\epsilon}_{23}}{2\sqrt{|a_{2}||a_{3}|}}\sqrt{\frac{b_{2}}{b_{3}}}\frac{|a_{3}|}{|a_{2}|}\right)\approx\sqrt{\frac{|a_{2}|}{b_{2}}},\nonumber\\
\Psi_{02}&=&\sqrt{\frac{|a_{3}|}{b_{3}}}
\left(1-\frac{\widetilde{\epsilon}_{13}}{2\sqrt{|a_{3}||a_{1}|}}\sqrt{\frac{b_{3}}{b_{1}}}\frac{|a_{1}|}{|a_{3}|}
-\frac{\widetilde{\epsilon}_{23}}{2\sqrt{|a_{2}||a_{3}|}}\sqrt{\frac{b_{3}}{b_{2}}}\frac{|a_{2}|}{|a_{3}|}\right)\approx\sqrt{\frac{|a_{3}|}{b_{3}}}.
\end{eqnarray}
That is, the effect of the weak interband coupling on the OP $\Psi_{1,2,3}$ at $T=0$ is not essential, and it can be described as perturbation. At $T\rightarrow T_{c}$, we have $\Psi_{01,02,03}\rightarrow 0$, then, the following approximation can be proposed:
\begin{eqnarray}\label{4.5}
 &&\Psi_{01}^{2}=\big({-a_{1}a_{2}a_{3}-2\widetilde{\epsilon}_{12}\widetilde{\epsilon}_{13}\widetilde{\epsilon}_{23}
 +\widetilde{\epsilon}_{23}^{2}a_{1}+\widetilde{\epsilon}_{13}^{2}a_{2}+\widetilde{\epsilon}_{12}^{2}a_{3}}\big)/
 {b_{1}\big(a_{2}a_{3}-\widetilde{\epsilon}_{23}^{2}\big)},\nonumber\\
 &&\Psi_{02}^{2}=\big({-a_{1}a_{2}a_{3}-2\widetilde{\epsilon}_{12}\widetilde{\epsilon}_{13}\widetilde{\epsilon}_{23}
 +\widetilde{\epsilon}_{23}^{2}a_{1}+\widetilde{\epsilon}_{13}^{2}a_{2}+\widetilde{\epsilon}_{12}^{2}a_{3}}\big)/
 {b_{2}\big(a_{1}a_{3}-\widetilde{\epsilon}_{13}^{2}\big)},\nonumber\\
 &&\Psi_{03}^{2}=\big({-a_{1}a_{2}a_{3}-2\widetilde{\epsilon}_{12}\widetilde{\epsilon}_{13}\widetilde{\epsilon}_{23}
 +\widetilde{\epsilon}_{23}^{2}a_{1}+\widetilde{\epsilon}_{13}^{2}a_{2}+\widetilde{\epsilon}_{12}^{2}a_{3}}\big)/
 {b_{3}\big(a_{1}a_{2}-\widetilde{\epsilon}_{12}^{2}\big)}.
\end{eqnarray}
Thus, at high temperatures $T\gtrsim T_{c1},T_{c2},T_{c3}$, the values of the OP $\Psi_{01,02,03}$ are determined by the interband couplings $\epsilon_{ik}$, so that, if $\epsilon_{12}=\epsilon_{13}=\epsilon_{23}=0$, then $\Psi_{01,02,03}=0$.

Let us introduce the following notes:
\begin{equation}\label{4.6}
  \alpha_{1}\equiv a_{1}+3b_{1}\Psi_{01}^{2},\quad \alpha_{2}\equiv a_{2}+3b_{2}\Psi_{02}^{2},\quad\alpha_{3}\equiv a_{3}+3b_{3}\Psi_{03}^{2},
\end{equation}
then,
\begin{eqnarray}\label{4.7}
\alpha_{1,2,3}&=& a_{1,2,3}>0, \quad \mathrm{at}  \quad T=T_{c},\nonumber \\
\alpha_{1,2,3}&=&-2a_{1,2,3}=2|a_{1,2,3}|, \quad \mathrm{at} \quad T\ll T_{c1},T_{c2},T_{c3}.
\end{eqnarray}
The second formula is correct if the weak interband coupling $\epsilon^{2}\ll a_{1}a_{2}$ takes place only. Lagrange equations for Lagrangian (\ref{4.2}) are:
\begin{eqnarray}\label{4.8}
  \frac{\hbar^{2}}{4m_{1}}\widetilde{\partial}_{\mu}\widetilde{\partial}^{\mu}\phi_{1}
  +\frac{\hbar^{2}}{4}\widetilde{\eta}_{12}\widetilde{\partial}_{\mu}\widetilde{\partial}^{\mu}\phi_{2}
  +\frac{\hbar^{2}}{4}\widetilde{\eta}_{13}\widetilde{\partial}_{\mu}\widetilde{\partial}^{\mu}\phi_{3}
  +\alpha_{1}\phi_{1}+\widetilde{\epsilon}_{12}\phi_{2}+\widetilde{\epsilon}_{13}\phi_{3} &=& 0,\nonumber\\
 \frac{\hbar^{2}}{4m_{2}}\widetilde{\partial}_{\mu}\widetilde{\partial}^{\mu}\phi_{2}
  +\frac{\hbar^{2}}{4}\widetilde{\eta}_{12}\widetilde{\partial}_{\mu}\widetilde{\partial}^{\mu}\phi_{1}
  +\frac{\hbar^{2}}{4}\widetilde{\eta}_{23}\widetilde{\partial}_{\mu}\widetilde{\partial}^{\mu}\phi_{3}
  +\alpha_{2}\phi_{2}+\widetilde{\epsilon}_{12}\phi_{1}+\widetilde{\epsilon}_{23}\phi_{3} &=& 0,\nonumber\\
  \frac{\hbar^{2}}{4m_{3}}\widetilde{\partial}_{\mu}\widetilde{\partial}^{\mu}\phi_{3}
  +\frac{\hbar^{2}}{4}\widetilde{\eta}_{13}\widetilde{\partial}_{\mu}\widetilde{\partial}^{\mu}\phi_{1}
  +\frac{\hbar^{2}}{4}\widetilde{\eta}_{23}\widetilde{\partial}_{\mu}\widetilde{\partial}^{\mu}\phi_{2}
  +\alpha_{3}\phi_{2}+\widetilde{\epsilon}_{13}\phi_{1}+\widetilde{\epsilon}_{23}\phi_{3} &=& 0.
\end{eqnarray}
The fields $\phi_{1,2,3}$ can be written in the form of harmonic oscillations: $\phi_{1}=A\exp({-\ri q_{\mu}x^{\mu}})$, $\phi_{2}=B\exp({-\ri q_{\mu}x^{\mu}})$, $\phi_{3}=C\exp({-\ri q_{\mu}x^{\mu}})$, where $q_{\mu}x^{\mu}=\omega t-\mathbf{qr}$. Substituting them in equation~(\ref{4.8}), we obtain equations for the amplitudes $A,B,C$:
\begin{eqnarray}\label{4.9}
   A\left(\alpha_{1}-q_{\mu}q^{\mu}\frac{\hbar^{2}}{4m_{1}}\right)
  +B\left(\widetilde{\epsilon}_{12}-q_{\mu}q^{\mu}\frac{\hbar^{2}}{4}\widetilde{\eta}_{12}\right)
  +C\left(\widetilde{\epsilon}_{13}-q_{\mu}q^{\mu}\frac{\hbar^{2}}{4}\widetilde{\eta}_{13}\right)&=&0,\nonumber \\
  A\left(\widetilde{\epsilon}_{12}-q_{\mu}q^{\mu}\frac{\hbar^{2}}{4}\widetilde{\eta}_{12}\right)
  +B\left(\alpha_{2}-q_{\mu}q^{\mu}\frac{\hbar^{2}}{4m_{2}}\right)
  +C\left(\widetilde{\epsilon}_{23}-q_{\mu}q^{\mu}\frac{\hbar^{2}}{4}\widetilde{\eta}_{23}\right)&=&0,\nonumber \\
   A\left(\widetilde{\epsilon}_{13}-q_{\mu}q^{\mu}\frac{\hbar^{2}}{4}\widetilde{\eta}_{13}\right)
  +B\left(\widetilde{\epsilon}_{23}-q_{\mu}q^{\mu}\frac{\hbar^{2}}{4}\widetilde{\eta}_{23}\right)
  +C\left(\alpha_{3}-q_{\mu}q^{\mu}\frac{\hbar^{2}}{4m_{3}}\right)&=&0.
\end{eqnarray}
Equating the determinant of the system to zero (\ref{4.9}), we find the dispersion equation:
\begin{equation}\label{4.10}
\left(q_{\mu}q^{\mu}\right)^{3}a+\left(q_{\mu}q^{\mu}\right)^{2}b+\left(q_{\mu}q^{\mu}\right)c+d=0,
\end{equation}
where
\begin{eqnarray}\label{4.10a}
  a &=& \left(\frac{\hbar^{2}}{4}\right)^{3}\left(\frac{1}{m_{1}m_{2}m_{3}}+2\widetilde{\eta}_{12}\widetilde{\eta}_{13}\widetilde{\eta}_{23}
  -\frac{\widetilde{\eta}_{23}^{2}}{m_{1}}-\frac{\widetilde{\eta}_{13}^{2}}{m_{2}}-\frac{\widetilde{\eta}_{12}^{2}}{m_{3}}\right),\nonumber\\
  b &=& \left(\frac{\hbar^{2}}{4}\right)^{2}\bigg(-\frac{\alpha_{1}}{m_{2}m_{3}}-\frac{\alpha_{2}}{m_{1}m_{3}}-\frac{\alpha_{3}}{m_{2}m_{2}}
  -2\widetilde{\epsilon}_{12}\widetilde{\eta}_{13}\widetilde{\eta}_{23}-2\widetilde{\epsilon}_{13}\widetilde{\eta}_{12}\widetilde{\eta}_{23}
  -2\widetilde{\epsilon}_{23}\widetilde{\eta}_{12}\widetilde{\eta}_{13}\nonumber\\
  &+&\alpha_{1}\widetilde{\eta}_{23}^{2}+\alpha_{2}\widetilde{\eta}_{13}^{2}+\alpha_{3}\widetilde{\eta}_{12}^{2}
  +2\frac{\widetilde{\epsilon}_{23}\widetilde{\eta}_{23}}{m_{1}}+2\frac{\widetilde{\epsilon}_{13}\widetilde{\eta}_{13}}{m_{2}}
  +2\frac{\widetilde{\epsilon}_{12}\widetilde{\eta}_{12}}{m_{3}}\bigg),\\
  c &=& \frac{\hbar^{2}}{4}\bigg[\frac{\widetilde{\epsilon}_{23}^{2}-\alpha_{2}\alpha_{3}}{m_{1}}
  +\frac{\widetilde{\epsilon}_{13}^{2}-\alpha_{1}\alpha_{3}}{m_{2}}
  +\frac{\widetilde{\epsilon}_{12}^{2}-\alpha_{1}\alpha_{2}}{m_{3}}\nonumber\\
  &-&2\widetilde{\eta}_{12}(\widetilde{\epsilon}_{13}\widetilde{\epsilon}_{23}-\alpha_{3}\widetilde{\epsilon}_{12})
  -2\widetilde{\eta}_{13}(\widetilde{\epsilon}_{12}\widetilde{\epsilon}_{23}-\alpha_{2}\widetilde{\epsilon}_{13})
  -2\widetilde{\eta}_{23}(\widetilde{\epsilon}_{12}\widetilde{\epsilon}_{13}-\alpha_{1}\widetilde{\epsilon}_{23})\bigg],\nonumber\\
  d &=& \alpha_{1}\alpha_{2}\alpha_{3}+2\widetilde{\epsilon}_{12}\widetilde{\epsilon}_{13}\widetilde{\epsilon}_{23}
  -\alpha_{1}\widetilde{\epsilon}_{23}^{2}-\alpha_{2}\widetilde{\epsilon}_{13}^{2}-\alpha_{3}\widetilde{\epsilon}_{12}^{2}.\nonumber
\end{eqnarray}
It should be noted that $d(T_{c})=0$ according to equations~(\ref{1.3}), \eqref{3.7}, \eqref{4.6}, and \eqref{4.7}. Hence, we have the corresponding dispersion relations at a critical temperature:
\begin{eqnarray}
  q_{\mu}q^{\mu}(T_{c})&=&0\label{4.11a},\\
  q_{\mu}q^{\mu}(T_{c})&=&\frac{-b\pm\sqrt{b^{2}-4ac}}{2a}>0.
  \label{4.11b}
\end{eqnarray}
We can see that Higgs mode splits to three branches. For the first mode (\ref{4.11a}), the energy gap (the mass of Higgs boson) vanishes at the critical temperature, as in single-band superconductors, and amplitudes of these modes relate as $A=B=C$. At the same time, the energy gap of the second and third modes~(\ref{4.11b}) does not vanish at the critical temperature. Thus, let us consider a case of symmetrical bands $\alpha_{1}=\alpha_{2}=\alpha_{3}\equiv\alpha$,  $\widetilde{\epsilon}_{1}=\widetilde{\epsilon}_{2}=\widetilde{\epsilon}_{3}\equiv\widetilde{\epsilon}$,  $m_{1}=m_{2}=m_{3}\equiv m$ and the drag effect is absent: $\eta_{12}=\eta_{13}=\eta_{23}=0$, then, the massive modes have the same spectrum ($b^{2}-4ac=0$):
\begin{equation}\label{4.12}
    q_{\mu}q^{\mu}(T_{c})=-\frac{12}{\hbar^{2}}m\widetilde{\epsilon},
\end{equation}
where $\widetilde{\epsilon}<0$. Amplitudes of these modes relate as, for example, $A=-C$, $B=0$ and $A=C$, $B=-(A+C)$, accordingly. 

In~\cite{grig2} it was demonstrated how the energy gap $\hbar\omega_{0}$ ($\mathbf{q}=0$) is related to the coherence length $\xi$: $\xi^{2}={2\upsilon^{2}}/{\omega^{2}_{0}}$ (or from the uncertainty principle: $\hbar\omega_{0}{\xi}/{\upsilon}\sim\hbar\Rightarrow\xi\sim{\upsilon}/{\omega_{0}}$, since the energy of Higgs mode plays the role of the uncertainty of energy in a superconductor). Thus, there are three coherence lengths according to the branches (\ref{4.11a}) and (\ref{4.11b}). For example, for the symmetrical bands without the drag effect, we obtain at $T=T_{c}$:
\begin{eqnarray}
  \xi_{1}^{2} &=& \infty, \label{4.13a}\\
  \xi_{2}^{2}=\xi_{3}^{2} &=& \frac{\hbar^{2}}{6m|\widetilde{\epsilon}|}< \infty. \label{4.13b}
\end{eqnarray}
The first coherence length diverges at $T=T_{c}$. On the contrary, the second and third lengths remain finite and they  vary only a little with temperature.

Thus, Higgs modes are oscillations of SC densities $n_{\mathrm{s}i}=2|\Psi_{i}|^{2}$. At the same time, the normal density must oscillate in anti-phase, so that the total density is constant $n=n_{\mathrm{s}}+n_{\mathrm{n}}=\mathrm{const}$,  hence,  $n_{\mathrm{s}}\mathbf{v}_{s}+n_{\mathrm{n}}\mathbf{v}_{n}=0$. Then, in order to change SC density, one Cooper pair must be broken as minimum, that is the energy of order of $2|\Delta|$ must be spent. Thus, in~\cite{grig2} it was demonstrated that in single-band superconductors $q_{\mu}q^{\mu}=4|\Delta|^{2}$.  \emph{Thus, to excite any Higgs mode at $T=T_{c}$ it is not necessary to spend this threshold energy, since $|\Delta(T_{c})|=0$. However, for the second and third branches} --- equation~(\ref{4.11b}) or equation~(\ref{4.12}), \emph{we have $q_{\mu}q^{\mu}(T_{c})\neq 0$, which is a nonphysical property}. Thus, we must assume equation~(\ref{3.9}), then from equation~(\ref{4.10a}) we can see that $a=b=0$. Hence, the anti-phase Higgs modes are absent, and the common mode oscillations with zero energy gap at $T=T_{c}$ remain only:
\begin{eqnarray}\label{4.14}
    &&q_{\mu}q^{\mu}=-\frac{d}{c}
   =\frac{4}{\hbar^{2}}\Bigg[\big({\alpha_{1}\alpha_{2}\alpha_{3}+2\widetilde{\epsilon}_{12}\widetilde{\epsilon}_{13}\widetilde{\epsilon}_{23}
   -\alpha_{1}\widetilde{\epsilon}_{23}^{2}-\alpha_{2}\widetilde{\epsilon}_{13}^{2}-\alpha_{3}\widetilde{\epsilon}_{12}^{2}}\big)
/
  \left(\frac{\alpha_{2}\alpha_{3}-\widetilde{\epsilon}_{23}^{2}}{m_{1}} \right.\nonumber\\
  	&&+\left.\frac{\alpha_{1}\alpha_{3}-\widetilde{\epsilon}_{13}^{2}}{m_{2}}
  +\frac{\alpha_{1}\alpha_{2}-\widetilde{\epsilon}_{12}^{2}}{m_{3}}
  -2\frac{\alpha_{3}\widetilde{\epsilon}_{12}-\widetilde{\epsilon}_{13}\widetilde{\epsilon}_{23}}{\sqrt{m_{1}m_{2}}}
  -2\frac{\alpha_{2}\widetilde{\epsilon}_{13}-\widetilde{\epsilon}_{12}\widetilde{\epsilon}_{23}}{\sqrt{m_{1}m_{3}}}
  -2\frac{\alpha_{1}\widetilde{\epsilon}_{23}-\widetilde{\epsilon}_{12}\widetilde{\epsilon}_{13}}{\sqrt{m_{2}m_{3}}}\right)\Bigg].\qquad
\end{eqnarray}
Respectively, there is only one coherence length $\xi(T)$:
\begin{equation}\label{4.15}
  \xi^{2}=\frac{2\upsilon^{2}}{\omega^{2}_{0}}=-\frac{2c}{d}.
\end{equation}
For symmetrical bands, we have the following dispersion law for the Higgs mode:
\begin{equation}\label{4.16}
  q_{\mu}q^{\mu}=\frac{4m}{3\hbar^{2}}(\alpha-2|\widetilde{\epsilon}|),
\end{equation}
whose energy gap vanishes at $T=T_{c}$: from equations~(\ref{1.4}) and (\ref{4.7}) we have $\alpha(T_{c})=a(T_{c})=2|\widetilde{\epsilon}|$, hence, $q_{\mu}q^{\mu}(T_{c})=0$.  The corresponding coherence length is:
\begin{equation}\label{4.17}
  \xi^{2}=\frac{3\hbar^{2}}{2m}\frac{1}{\left|\alpha-2|\widetilde{\epsilon}|\right|},
\end{equation}
so that $\xi(T_{c})=\infty$.

We could see from the properties of Higgs modes that the  \emph{existence of several coherence lengthes with corresponding properties is incompatible with the second-order phase transition}. Then, if equation~(\ref{3.9}) takes place, then Leggett modes are absent, and the common mode oscillations with acoustic spectrum~(\ref{3.8}) remain only. \emph{Thus, as in single-band superconductors, in three-band superconductors the common mode oscillations exist only. The anti-phase Goldstone mode (i.e., Leggett modes) and the anti-phase Higgs modes are absent, which ensures only one coherence length $\xi(T)$ diverging at $T=T_{c}$}. At the same time, the Goldstone mode is accompanied by current. Therefore, the gauge field $\widetilde{A}_{\mu}$ absorbs the Goldstone boson $\theta$, as in single-band superconductors, i.e., Anderson-Higgs mechanism takes place~\cite{grig2}. The condition (\ref{3.9}) generalizes the condition obtained in~\cite{grig1,grig3} for two-band superconductors, which prohibits type 1.5 superconductors.

Let us consider the regime of almost independent condensates in each band. This means: 1)~temperature must be low, i.e., $T\ll T_{c1},T_{c2},T_{c3}$, 2)~the weak interband coupling $\epsilon_{ik}^{2}\ll a_{i}a_{k}$ must take place. Using equation~(\ref{4.4}), the energy gap $\hbar\omega_{0}$ ($\mathbf{q}=0$) of Higgs mode (\ref{4.14}) can be reduced to a form:
\begin{eqnarray}\label{4.18}
  (\hbar\omega_{0})^{2}&=&4\upsilon^{2}
  \frac{\alpha_{1}\alpha_{2}\alpha_{3}}{({\alpha_{2}\alpha_{3}}/{m_{1}})+({\alpha_{1}\alpha_{3}}/{m_{2})+({\alpha_{1}\alpha_{2}}/{m_{3}})}}\nonumber\\
  &=&8\upsilon^{2}\frac{|a_{1}||a_{2}||a_{3}|}{({|a_{2}||a_{3}|}/{m_{1}})+({|a_{1}||a_{3}|}/{m_{2}})+({|a_{1}||a_{2}|}/{m_{3}})}\nonumber\\
  &=&\frac{8}{3}\upsilon^{2}\left[\frac{\sqrt{a_{1}^{2}|a_{2}||a_{3}|b_{2}b_{3}}}
  {({|a_{2}||a_{3}|}/{m_{1}})+({|a_{1}||a_{3}|}/{m_{2}})+({|a_{1}||a_{2}|}/{m_{3}})}\Psi_{02}\Psi_{03}\right.\nonumber\\
  &+&\frac{\sqrt{a_{2}^{2}|a_{1}||a_{3}|b_{1}b_{3}}}
  {({|a_{2}||a_{3}|}/{m_{1}})+({|a_{1}||a_{3}|}/{m_{2}})+({|a_{1}||a_{2}|}/{m_{3}})}\Psi_{01}\Psi_{03}\nonumber\\
  &+&\left.\frac{\sqrt{a_{3}^{2}|a_{1}||a_{2}|b_{1}b_{2}}}
  {({|a_{2}||a_{3}|}/{m_{1}})+({|a_{1}||a_{3}|}/{m_{2}})+({|a_{1}||a_{2}|}/{m_{3}})}\Psi_{01}\Psi_{02}\right].
\end{eqnarray}
Then, multipliers before $\Psi_{0i}\Psi_{0k}$ depend on temperature very weakly, and this energy is symmetrical with respect to the bands. Using the relationship between the ``wave function'' of Cooper pairs $\Psi$ and the energy gap $\Delta$~\cite{grig2,sad1,levit}:
\begin{equation}\label{4.19}
   \Psi_{i}=\frac{\left[14\zeta(3)n_{i}\right]^{1/2}}{4\piup T_{ci}}\Delta_{i},
\end{equation}
where $n_{i}={k_{Fi}^{3}}/{3\piup^{2}}$ is electron density for a band $i$. Then, we can see that $(\hbar\omega_{0})^{2}\propto|\Delta_{i}||\Delta_{k}|$, and we can assume:
\begin{equation}\label{4.20}
  (\hbar\omega_{0})^{2}=\chi_{12}\Delta_{01}\Delta_{02}+\chi_{13}\Delta_{01}\Delta_{03}+\chi_{23}\Delta_{02}\Delta_{03},
\end{equation}
where $\chi_{ik}=\mathrm{const}$ (dimensionless) are such that in superconductor with symmetrical $m_{1}=m_{2}=m_{3}$,  $n_{1}=n_{2}=n_{3}$, $a_{1}=a_{2}=a_{3}$, $b_{1}=b_{2}=b_{3}$,  $T_{c1}=T_{c2}=T_{c3}$ $\Rightarrow$ $\Delta_{1}=\Delta_{2}=\Delta_{3}$ and almost independent bands (i.e., $\epsilon^{2}_{ik}\ll a_{i}a_{k}$ at $T\ll T_{c1},T_{c2},T_{c3}$), we should have $\upsilon={\upsilon_{\rm F}}/{3}$, since in single-band superconductors we have $\upsilon={\upsilon_{\rm F}}/{\sqrt{3}}$ and we can determine the ``dielectric permittivity'' as $\varepsilon={c^{2}}/{\upsilon^{2}}={c^{2}}/({\upsilon_{\rm F}^{2}/3})$~\cite{grig2}, then a ``mixture'' of three superconductors is equivalent to three parallel dielectrics (capacitors), then the total permittivity is $\varepsilon=\varepsilon_{1}+\varepsilon_{2}+\varepsilon_{3}={3c^{2}}/\left({\upsilon_{\rm F}^{2}/3}\right)$; hence, we obtain for the ``mixture'': $\upsilon={\upsilon_{\rm F}}/{3}$. The coefficients $a_{i}$, $b_{i}$ are~\cite{sad2}:
\begin{equation}\label{4.21}
  a_{i}=\frac{6\piup^{2}T_{ci}}{7\zeta(3)\varepsilon_{Fi}}\left(T-T_{ci}\right),\quad b_{i}=\frac{6\piup^{2}T_{ci}}{7\zeta(3)\varepsilon_{Fi}}\frac{T_{ci}}{n_{i}}.
\end{equation}
Let us suppose that $\chi_{12}\approx\chi_{13}\approx\chi_{23}\equiv\chi=\mathrm{const}$, and consider symmetrical bands (in particular $\upsilon_{\rm F 1}=\upsilon_{\rm F 2}=\upsilon_{\rm F 3}\equiv\upsilon_{\rm F }$).
Substituting equations~(\ref{4.18}), \eqref{4.19} and (\ref{4.21}) in equation~(\ref{4.20}) we obtain:
\begin{equation}\label{4.22}
  \upsilon^{2}=\frac{3\chi}{4}\upsilon_{\rm F }^{2}\Rightarrow\chi=\frac{4}{27}.
\end{equation}
For the material with different bands at $T\ll T_{c1},T_{c2},T_{c3}$ we can obtain the following approximation:
\begin{equation}\label{4.23}
  \upsilon^{2}\approx\frac{1}{9}\left(\frac{T_{c2}T_{c3}}{T_{c1}}\upsilon_{\rm F 1}^{2}
  +\frac{T_{c1}T_{c3}}{T_{c2}}\upsilon_{\rm F 2}^{2}+\frac{T_{c1}T_{c2}}{T_{c3}}\upsilon_{\rm F 3}^{2}\right)\frac{1}{T_{c1}+T_{c2}+T_{c3}}.
\end{equation}
Thus, the speed of ``light''  $\upsilon$ is of the order of Fermi speeds $\upsilon_{\rm F 1}$, $\upsilon_{\rm F 2}$, $\upsilon_{\rm F 3}$ in the corresponding bands, as in single-band superconductors, where $\upsilon={\upsilon_{\rm F }}/{\sqrt{3}}$~\cite{grig2}.

\section{Results}\label{results}

In this work we investigate equilibrium states, magnetic response and the normal oscillations of internal degrees of freedom of three-band superconductors with the accounting of the terms of the ``drag'' effect $\eta_{ik}\left\{D_{\mu}\Psi_{i}(D^{\mu}\Psi_{k})^{+}+(D^{\mu}\Psi_{i})^{+}D_{\mu}\Psi_{k}\right\}$. Our results are as follows:

1) The obtained equation for critical temperature (\ref{1.3}) demonstrates that $T_{c}$ depends on the signs of the coefficients of internal proximity effect ($\epsilon_{ik}<0$ for attractive interband interaction, $\epsilon_{ik}>0$ for repulsive interband interaction): $T_{c}(\epsilon_{12}\epsilon_{13}\epsilon_{23}<0)>T_{c}(\epsilon_{12}\epsilon_{13}\epsilon_{23}>0)$.  As in two-band systems, the effect of interband coupling is nonperturbative: the application of the weak interband coupling washes out all OP up to a new critical temperature, as illustrated in figure~{\ref{Fig2}}. The magnetic penetration depth is determined with SC densities in each band, although the drag terms renormalize the carrier masses see equation~(\ref{1.9}).

2) Due to the internal proximity effect, the Goldstone mode splits into three branches: common mode oscillations with the acoustic spectrum, and the oscillations of the relative phases $\theta_{i}-\theta_{k}$ between SC condensates with an energy gap in the spectrum determined by interband couplings $\epsilon_{ik}$, which are analogous to the Leggett mode in two-band superconductors. The common mode oscillations are absorbed by the gauge field $A_{\mu}$. That is why oscillations are accompanied by current, as in single-band superconductors~\cite{grig2}. At the same time, the massive modes are not accompanied by current. Therefore, they ``survive''. If we assume that the coefficients of the drag effect $\eta_{ik}$ are such as in equation~(\ref{3.9}), then the Leggett modes are absent, and the common mode oscillations (\ref{3.8}) remain only.

3) Higgs oscillations also split into three branches. The energy gap of the common mode vanishes at critical temperature $T_{c}$, for the other two anti-phase modes their energy gaps do not vanish at $T_{c}$ and are determined by the interband couplings $\epsilon_{ik}$. The mass of Higgs mode is related to the coherence length~$\xi$. Hence, we obtain three coherence lengths accordingly. The first coherence length diverges at $T=T_{c}$, while on the contrary, the second and third lengths remain finite at all temperatures. The effect of the splitting of Goldstone and Higgs modes into three branches each takes place even at the infinitely small coefficients~$\epsilon_{ik}$. Thus, the effect of interband coupling $\epsilon\neq0$ is nonperturbative. As for Goldstone modes, if we assume that coefficients of the drag effect $\eta_{ik}$ are such as in equation~(\ref{3.9}), then the anti-phase Higgs modes are absent and the common mode oscillations (\ref{4.14}) with zero energy gap at $T=T_{c}$ remain only.

4) The excitation of one quant of Higgs oscillations requires the breaking of one Cooper pair as minimum, i.e., the energy of the order of $2|\Delta|$ must be spent. Hence, to excite any Higgs mode at $T=T_{c}$ it is not necessary to spend this threshold energy. In  three-band superconductors for anti-phase Higgs modes, we have a nonphysical property $q_{\mu}q^{\mu}(T_{c})\neq 0$. As and for Goldstone modes, if we assume that coefficients of the drag effect $\eta_{ik}$ are the same as in equation~(\ref{3.9}), then the anti-phase Higgs modes are absent and the common mode oscillations with zero energy gap at $T=T_{c}$ remain only. \emph{Thus, as in single-band superconductors, in three-band superconductors the common mode oscillations exist only. The anti-phase Goldstone mode (i.e., Leggett modes) and the anti-phase Higgs modes are absent, which ensures only single coherence length $\xi(T)$ diverging at $T=T_{c}$}.


5) The square of the energy gap of Higgs mode in three-band superconductors can be represent in the form of a sum of products of gaps $\Delta_{0i}\Delta_{0k}$ see equation~(\ref{4.20}), which is similar to two-band superconductors~\cite{grig3}, and it differs from the mass of Higgs mode in single-band superconductors: $\hbar\omega_{0}=2|\Delta|$, where this mode exists in the free quasiparticle continuum. On the contrary, in two-band superconductors and in three-band superconductors it can be $\sqrt{|\Delta_{i}||\Delta_{k}|}<2\min(|\Delta_{1}|,|\Delta_{2}|,|\Delta_{3}|)$, then the Higgs mode becomes stable. The speed of ``light''  $\upsilon$ is of the order of Fermi velocities in each band $\upsilon_{\rm F 1},\upsilon_{\rm F 2},\upsilon_{\rm F 3}$ and depends on the single-band ``critical'' temperatures $T_{c1},T_{c2},T_{c3}$ see equation~(\ref{4.23}).

6) Unlike the two-band systems, the Higgs modes and the Goldstone modes can be hybridized at $\epsilon_{12}\epsilon_{13}\epsilon_{23}>0$. For the case $\epsilon_{12}\epsilon_{13}\epsilon_{23}<0$, the hybridization is absent. All previous results were obtained in the approximation of splitting of the correlation between amplitude and phase oscillations.
\newpage

\section*{Acknowledgements}
This research
was supported by grant of National Research Foundation of Ukraine ``Models
of nonequilibrium processes in colloidal systems'' 2020.02/0220, by theme
grant of Department of physics and astronomy of NAS of Ukraine:
``Noise-induced dynamics and correlations in nonequilibrium systems''
0120U101347 and by grant of Simons Foundation.

\appendix
\section{Some symmetric 3HDM potentials}\label{symm}

Following to~\cite{keus} a scalar 3HDM potential symmetric under a group $G$ can be written as
\begin{equation}\label{A1}
  V=V_{0}+V_{G},
\end{equation}
where
\begin{eqnarray}\label{A2}
    V_{0}&=&\sum_{i=1}^{3}a_{i}\left|\Psi_{i}\right|^{2}+\frac{b_{i}}{2}\left|\Psi_{i}\right|^{4}\nonumber\\
    &+&b_{12}|\Psi_{1}|^{2}|\Psi_{2}|^{2}+b_{13}|\Psi_{1}|^{2}|\Psi_{3}|^{2}+b_{23}|\Psi_{2}|^{2}|\Psi_{3}|^{2}\nonumber\\
    &+&b_{12}'(\Psi_{1}^{+}\Psi_{2})(\Psi_{2}^{+}\Psi_{1})+b_{13}'(\Psi_{1}^{+}\Psi_{3})(\Psi_{3}^{+}\Psi_{1})+
    b_{23}'(\Psi_{2}^{+}\Psi_{3})(\Psi_{3}^{+}\Psi_{2})
\end{eqnarray}
is invariant under the most general $U(1)\times U(1)$ gauge transformation and $U_{G}$ is a collection of extra terms
ensuring the symmetry group $G$. The $U(1)\times U(1)$ group is generated by
\begin{equation}\label{A3}
  \left(
     \begin{array}{ccc}
       \re^{-\ri \alpha} & 0 & 0 \\
       0 & \re^{\ri\alpha} & 0 \\
       0 & 0 & 1 \\
     \end{array}
   \right)
   \left(
     \begin{array}{ccc}
       \re^{-2\ri\beta/3} & 0 & 0 \\
       0 & \re^{\ri\beta/3} & 0 \\
       0 & 0 & \re^{\ri\beta/3} \\
     \end{array}
   \right).
\end{equation}
However, in the present work we use the minimum model, where
$b_{ik}=b_{ik}'=0$. A potential symmetric under the $U(1)$ group is
\begin{equation}\label{A4}
  V_{U(1)}=V_{0}+\lambda_{123}\left[(\Psi_{1}^{+}\Psi_{3})(\Psi_{2}^{+}\Psi_{3})+(\Psi_{1}\Psi_{3}^{+})(\Psi_{2}\Psi_{3}^{+})\right].
\end{equation}
The $U(1)$ group is generated by
\begin{equation}\label{A5}
  \left(
     \begin{array}{ccc}
       \re^{-\ri \alpha} & 0 & 0 \\
       0 & \re^{\ri\alpha} & 0 \\
       0 & 0 & 1 \\
     \end{array}
   \right).
\end{equation}
A potential symmetric under the $U(1)\times Z_{2}$ group is
\begin{equation}\label{A6}
  V_{U(1)\times Z_{2}}=V_{0}+\lambda_{23}\left[(\Psi_{2}^{+}\Psi_{3})^{2}+(\Psi_{2}\Psi_{3}^{+})^{2}\right].
\end{equation}
The $U(1)\times Z_{2}$ group is generated by
\begin{equation}\label{A7}
  \left(
     \begin{array}{ccc}
       \re^{-2\ri\beta/3} & 0 & 0 \\
       0 & \re^{\ri\beta/3} & 0 \\
       0 & 0 & \re^{\ri\beta/3} \\
     \end{array}
   \right)
   \left(
     \begin{array}{ccc}
       -1 & 0 & 0 \\
       0 & -1 & 0 \\
       0 & 0 & 1 \\
     \end{array}
   \right).
\end{equation}
A potential symmetric under the $Z_{2}$ group is
\begin{eqnarray}\label{A8}
  V_{Z_{2}}&=&V_{0}+\epsilon_{12}\left[\Psi_{1}^{+}\Psi_{2}+\Psi_{1}\Psi_{2}^{+}\right]
  +\lambda_{12}\left[\left(\Psi_{1}^{+}\Psi_{2}\right)^{2}+\left(\Psi_{1}\Psi_{2}^{+}\right)^{2}\right]\nonumber\\
  &+&\lambda_{13}\left[\left(\Psi_{1}^{+}\Psi_{3}\right)^{2}+\left(\Psi_{1}\Psi_{3}^{+}\right)^{2}\right]
  +\lambda_{23}\left[\left(\Psi_{2}^{+}\Psi_{3}\right)^{2}+\left(\Psi_{2}\Psi_{3}^{+}\right)^{2}\right].
\end{eqnarray}
The $Z_{2}$ group is generated by
\begin{equation}\label{A9}
   \left(
     \begin{array}{ccc}
       -1 & 0 & 0 \\
       0 & -1 & 0 \\
       0 & 0 & 1 \\
     \end{array}
   \right).
\end{equation}


\begin{thebibliography}{99}
\bibitem{grig2} Grigorishin~K.~V., J. Low Temp. Phys., 2021 \textbf{203},  262, \doi{10.1007/s10909-021-02580-0}.
\bibitem{ars} Arseev~P.~I., Loiko~S.~O., Fedorov~N.~K., Phys. Usp., 2006, \textbf{49}, 1,\\ \doi{10.1070/PU2006v049n01ABEH002577}.
\bibitem{volovik} Volovik~G.~E., Zubkov~M.~A., J. Low Temp. Phys., 2014, \textbf{175}, 486, \doi{10.1007/s10909-013-0905-7}.

\bibitem{asker7} Askerzade~I.~N., Unconventional Superconductors: Anisotropy and Multiband Effects, Springer, Berlin, 2012.
\bibitem{grig1} Grigorishin~K.~V., Phys. Lett. A, 2016, \textbf{380}, 1781, \doi{10.1016/j.physleta.2016.03.023}.
\bibitem{asker2} Askerzade~I.~N., Phys. Usp., 2006, \textbf{49}, 1003, \doi{10.1070/PU2006v049n10ABEH006055}.
\bibitem{yerin1} Yerin~Y.~S., Omelyanchouk~A.~N., Low Temp. Phys., 2007, \textbf{33}, 401, \doi{10.1063/1.2737547}.
\bibitem{grig3} Grigorishin~K.~V., J. Low Temp. Phys., 2022, \textbf{206},  360, \doi{10.1007/s10909-022-02668-1}. 	
\bibitem{pono1} Ponomarev~Ya.~G., Kuzmichev~S.~A., Mikheev~M.~G., Sudakova~M.~V., Tchesnokov~S.~N., Timergaleev~N.~Z., Yarigin~A.~V., Maksimov~E.~G., Krasnosvobodtsev~S.~I., Varlashkin~A.~V., Hein~M.~A., M\"uller~G., Piel~H., Sevastyanova~L.~G., Kravchenko~O.~V.,  Burdina~K.~P., Bulychev~B.~M., Solid State Commun., 2004, \textbf{129}, 85,\\ \doi{10.1016/j.ssc.2003.09.024}.
\bibitem{pono2} Ponomarev~Ya.~G., Kuzmichev~S.~A., Mikheev~M.~G., Sudakova~M.~V., Tchesnokov~S.~N., Van~Hoai~H., Bulychev~B.~M., Maksimov~E.~G., Krasnosvobodtsev~S.~I., JETP Lett., 2007, \textbf{85}, 46,\\ \doi{10.1134/S0021364007010092}.
\bibitem{kuzm1}  Kuzmicheva~T.~E., Kuzmichev~S.~A., Morozov~I.~V., Wurmehl~S., B\"uchner~B., JETP Lett., 2020, \textbf{111}, 350,\\ \doi{10.1134/S002136402006003X}.
\bibitem{kuzm2}  Kuzmicheva~T.~E., Kuzmichev~S.~A., JETP Lett., 2021, \textbf{114}, 630, \doi{10.1134/S0021364021220070}.
\bibitem{kord}  Kordyuk~A.~A., Zabolotnyy~V.~B., Evtushinsky~D.~V., Yaresko~A.~N., B\"uchner~B., Borisenko~S.~V.,\\  J.~Supercond.~Novel~Magn., 2013, \textbf{26}, 2837, \doi{10.1007/s10948-013-2210-8}.
\bibitem{scaff} Scaffidi~T., Ph.D. Thesis, Merton College
University of Oxford, 2016.
\bibitem{tanaka} Tanaka~Y., Yanagisawa~T., J. Phys. Soc. Jpn., 2010, \textbf{79}, 114706, \doi{10.1143/JPSJ.79.114706}.
\bibitem{stanev1} Stanev~V., Te\v{s}anovi\'{c}~Z., Phys. Rev. B, 2010, \textbf{81}, 134522, \doi{10.1103/PhysRevB.81.134522}.
\bibitem{stanev2} Stanev~V., Phys. Rev. B, 2012, \textbf{85}, 174520, \doi{10.1103/PhysRevB.85.174520}.
\bibitem{stanev3} Stanev~V., Supercond. Sci. Technol., 2015, \textbf{28}, 014006, \doi{10.1088/0953-2048/28/1/014006}.
\bibitem{babaev2} Bojesen~T.~A., Babaev~E., Sudb\o{} A., Phys. Rev. B, 2013, \textbf{88}, 220511(R), \doi{10.1103/PhysRevB.88.220511}.
\bibitem{maiti} Maiti~S., Chubukov A. V., Phys. Rev. B, 2013, \textbf{87}, 144511, \doi{10.1103/PhysRevB.87.144511}.
\bibitem{wilson} Wilson~B.~J., Das~M.~P., J. Phys.: Condens. Matter, 2013, \textbf{25}, 425702, \doi{10.1088/0953-8984/25/42/425702}.
\bibitem{dias} Dias~R.~G., Marques~A.~M., Supercond. Sci. Technol., 2011, \textbf{24}, 085009, \doi{10.1088/0953-2048/24/8/085009}.
\bibitem{yerin3} Yerin~Y., Omelyanchouk~A., Drechsler~S.~L., Efremov~D.~V., van~den~Brink~J., Phys. Rev. B, 2017, \textbf{96}, 144513,\\ \doi{10.1103/PhysRevB.96.144513}.
\bibitem{lin1} Lin~S., Hu~X., Phys. Rev. Lett., 2012, \textbf{108}, 177005, \doi{10.1103/PhysRevLett.108.177005}.
\bibitem{kobay} Kobayashi~K., Machida~M., Ota~Y., Nori~F., Phys. Rev. B, 2013, \textbf{88}, 224516, \doi{10.1103/PhysRevB.88.224516}.
\bibitem{yerin4} Yerin~Y., Drechsler~S.~L., Phys. Rev. B, 2021, \textbf{104}, 014518, \doi{10.1103/PhysRevB.104.014518}.
\bibitem{lin2} Lin~S., Hu~X., New J. Phys., 2012, \textbf{14}, 063021, \doi{10.1088/1367-2630/14/6/063021}.
\bibitem{babaev3} Garaud~J., Carlstr\"om~J., Babaev~E., Phys. Rev. Lett., 2011, \textbf{107}, 197001, \doi{10.1103/PhysRevLett.107.197001}.
\bibitem{babaev1} Carlstr\"om~J., Garaud~J., Babaev~E., Phys. Rev. B, 2011, \textbf{84}, 134518, \doi{10.1103/PhysRevB.84.134518}.
\bibitem{hu} Hu~X., Wang~Z., Phys. Rev. B, 2012, \textbf{85}, 064516, \doi{10.1103/PhysRevB.85.064516}.
\bibitem{huang} Huang~Z., Hu~X., Appl. Phys. Lett., 2014, \textbf{104}, 162602, \doi{10.1063/1.4872261}.
\bibitem{asker3} Askerzade~I., Matrasulov~D., Salati~M., J. Supercond. Novel Magn., 2022, \textbf{35}, 2749, \doi{10.1007/s10948-022-06343-0}.
\bibitem{yerin2} Yerin~Y.~S., Omelyanchouk~A.~N., Low Temp. Phys., 2014, \textbf{40}, 943, \doi{10.1063/1.4897416}.
\bibitem{yerin5} Yerin~Y.~S., Kiyko~A.~S., Omelyanchouk~A.~N.,  Il'ichev~E., Low Temp. Phys., 2015, \textbf{41}, 885,\\ \doi{10.1063/1.4935255}.
\bibitem{hase} Yanagisawa~T., Hase~I., J. Phys. Soc. Jpn., 2013, \textbf{82}, 124704, \doi{10.7566/JPSJ.82.124704}.
\bibitem{keus} Keus~V., King~S.~F., Moretti~S., J. High Energy Phys., 2014, \textbf{2014},  52, \doi{10.1007/JHEP01(2014)052}.
\bibitem{pal2} Kochorbe~F.~G., Palistrant~M.~E., Physica C, 1998, \textbf{298}, 217, \doi{10.1016/S0921-4534(98)00004-5}.
\bibitem{pal1} Palistrant~V.~A., Theor. Math. Phys., 1993, \textbf{95}, 432, \doi{10.1007/BF01015898}.
\bibitem{ota} Ota~Y., Machida~M., Koyama~T., Aoki~H., Phys. Rev. B, 2011, \textbf{83}, 060507(R), \doi{10.1103/PhysRevB.83.060507}.
\bibitem{sad1} Sadovskii~M.~V., Diagrammatics: Lectures on Selected Problems in Condensed Matter Theory, World Scientific, 2006.
\bibitem{levit} Levitov~L.~S., Shitov~A.~V., Green's Functions. Problems and Solutions, Fizmatlit, Moscow, 2003, (in Russian).
\bibitem{sad2} Sadovskii~M.~V., Statistical Physics, De Gruyte, Berlin, 2012.

\end{thebibliography}

%
%
\newpage

\ukrainianpart

\title{Коллективні збудження у тризонному надпровіднику}
\author{К. В. Григоришин}
\address{ Інститут теоретичної фізики ім. М.М. Боголюбова НАН України, вул. Метрологічна 14-б, 03143 Київ, Україна}
%
%
%

\makeukrtitle

\begin{abstract}
\tolerance=3000%
Досліджено стани рівноваги, магнітний відгук і нормальні коливання внутрішніх ступенів вільності (моди Хіггса та моди Голдстоуна) тризонних надпровідників з урахуванням як внутрішнього ефекту близькості, так і ефекту ``захоплення'' (міжградієнтної взаємодії) в лагранжіані. Як мода Голдстоуна, так і мода Хіггса розщеплюються на три гілки кожна: синфазні коливання та дві моди протифазних коливань, що аналогіч\-ні моді Леггетта в двозонних надпровідниках. Показано, що друга і третя гілки є нефізичними, і їх можна усунути спеціальним підбором коефіцієнтів при членах ``захоплення'' в лагранжіані. У результаті тризонні надпровідники характеризуються лише однією довжиною когерентності. Отримано спектр синфазних коливань Хіггса. Глибина магнітного проникнення визначається густиною надпровідних елект\-ронів у кож\-ній зоні, однак міжградієнтна взаємодія перенормує маси носіїв.
\keywords лоренц-коваріантність, мода Хіггса, мода Голдстоуна, мода Леггетта, міжзонна взаємодія, ефект захоплення

\end{abstract}

\end{document}